\documentclass{article}

\usepackage{arxiv}

\usepackage[utf8]{inputenc} 
\usepackage[T1]{fontenc}    
\usepackage{hyperref}       
\usepackage{url}            
\usepackage{booktabs}       
\usepackage{amsfonts}       
\usepackage{nicefrac}       
\usepackage{microtype}      
\usepackage{lipsum}
\usepackage{graphicx}
\graphicspath{ {./images/} }
\usepackage{threeparttable}

\usepackage{subcaption}

\usepackage{amsmath}
\usepackage{amssymb}
\usepackage{mathtools}
\usepackage{bm}

\usepackage{array}
\usepackage{booktabs}
\usepackage{multirow}

\usepackage{hyperref}

\usepackage[numbers]{natbib}

\usepackage{algorithm}
\usepackage{algpseudocode}

\usepackage{xcolor}
\usepackage{enumitem}
\usepackage{siunitx}
\usepackage{csquotes}

\title{A High Efficient and Scalable Obstacle-Avoiding VLSI Global Routing Flow}

\author{
 Junhao Guo \\
  School of Integrated Circuits\\
  Sun Yat-sen University\\
  \texttt{guojh73@mail2.sysu.edu.cn} \\
   \And
 Hongxin Kong \\
  Advanced Micro Devices, Inc.\\
  \texttt{hongxink@amd.com} \\
  \And
 Lang Feng \thanks{Corresponding author} \\
  School of Integrated Circuits\\
  Sun Yat-sen University\\
  \texttt{fenglang3@mail.sysu.edu.cn} \\
}

\begin{document}
\maketitle
\begin{abstract}
Routing is a crucial step in the VLSI design flow. With advancements in manufacturing technology, more constraints have emerged in design rules, particularly regarding obstacles during routing, leading to increased routing complexity. Unfortunately, many global routers struggle to generate efficient obstacle-free solutions due to the lack of scalable obstacle-avoiding tree generation methods and the capability to handle modern designs with complex obstacles and nets.
In this work, we propose an efficient obstacle-aware global routing flow for VLSI designs with obstacles. The flow includes a rule-based obstacle-avoiding rectilinear Steiner minimal tree (OARSMT) algorithm during the tree generation phase. This algorithm is both scalable and fast, providing tree topologies avoiding obstacles in the early stage globally. With its guidance, in the later stages, the OARSMT-guided and obstacle-aware sparse maze routing are proposed to further minimize obstacle violations and reduce overflow costs. Compared to previously advanced methods on the benchmark with obstacles, our approach successfully eliminates obstacle violations and reduces wirelength and overflow cost, while sacrificing only a limited number of via counts and runtime overhead. 
\end{abstract}

\keywords{Physical design, Global routing, Obstacle-avoiding rectilinear steiner minimal tree}

\section{Introduction}
\label{sec:intro}

Routing is a crucial step in the back-end design of very large-scale integrated circuits (VLSI). It consists of two main phases: global routing and detailed routing. Global routing focuses on generating routing paths for every net to optimize performance and minimize violations as much as possible. In contrast, detailed routing addresses and resolves any violations after the global routing phase.
The quality of global routing significantly impacts the workload of detailed routing and the overall performance of the circuit. Furthermore, improvements in global routing can often be achieved more easily with advanced algorithms. 
Consequently, both academia and industry have invested substantial efforts in developing fast, scalable, and robust global routing engines.

As the number of cells and nets in modern designs has increased dramatically in advanced technology nodes, more complicated obstacles have emerged in global routing to accommodate more complex functionality requirements or to maintain consistency with earlier design phases. These obstacles may arise from various sources, including macro cells, intellectual property (IP) blocks, 3D packaging, pre-routed nets, power networks, etc.
However, many global routers struggle to efficiently handle obstacle information and generate obstacle-free solutions at the global routing phase. High-efficiency obstacle-avoiding global routing has become more challenging and more necessary than ever before.

Global routing typically involves several steps, including tree generation, initial routing, and rip-up and reroute phases. Traditional routers that lack specialized methods for avoiding obstacles typically generate rectilinear Steiner minimal trees (RSMTs) for nets, which can be quickly obtained using established algorithms like FLUTE~\cite{Chu08}. 
Without full consideration of obstacles, numerous design-rule violations can occur in the solutions. Although these violations can be addressed in later stages, with more time-consuming algorithms like maze routing, the overall efficiency and effectiveness of the routing process can become limited. To address the obstacle-avoiding RSMT (OARSMT) problem, recent research has introduced several algorithms for single-net solutions, such as complex graph algorithms, machine learning, etc. With more complex obstacle distributions, the complexity can significantly increase, especially for high fan-out nets. Additionally, the number of nets has also risen dramatically in modern designs. These factors result in scalability problems for most OARSMT algorithms in practicality when applied to large-scale cases in global routing. As a result, most previous research primarily concentrate on the tree generation phase and lack a comprehensive obstacle-avoiding global routing flow. 

In this article, we present an efficient and scalable obstacle-avoiding global routing flow using the proposed rule-based OARSMT algorithm for initial routing, along with OARSMT-guided and obstacle-aware sparse maze routing for the rip-up and reroute process. These methods successfully generate valid solutions under different distributions of obstacles in large designs, demonstrating high efficiency. The main contributions of this article are as follows:

\begin{itemize}

    \item We propose a global routing flow using the proposed rule-based OARSMT algorithm to avoid obstacle violations at the early stage. 
    This is followed by two obstacle-aware rip-up and reroute approaches that ultimately generate obstacle-free solutions for all nets.

    \item We propose a rule-based OARSMT algorithm that utilizes multiple rule-based schemes to enhance the results from FLUTE and operates very quickly. With its high scalability, it can address most obstacle violations for all nets in a complete design during tree generation.
 
    \item An OARSMT-guided sparse maze routing is proposed, which is guided by the OARSMT structure generated in the initial routing. This significantly decreases obstacle violations and overflow during rip-up and reroute processes. Additionally, an obstacle-aware sparse maze routing technique is introduced to manage particularly hard-to-route nets.

    \item The proposed obstacle-avoiding global routing eliminates obstacle violations while achieving a 1.96\% improvement in wirelength and a 28.06\% reduction in overflow cost on ISPD24 benchmarks with various obstacles, with limited via count and runtime overhead, demonstrating high efficiency. Furthermore, the proposed OARSMT algorithm achieves significant speedups of $\sim$10x-2700x and $\sim$150x-5800x runtime compared to previously advanced OARSMT algorithms on randomized testcases and standard OARSMT benchmarks, respectively, with minimal wirelength overhead.

\end{itemize}

The remaining of this article is organized as follows: In Section II, a brief background and formulation on global routing and OARSMT is presented. In Section III, a brief introduction to our obstacle-avoiding global routing flow is proposed. In Section IV, the detailed OARSMT algorithm is described. In Section V, the details of our obstacle-avoiding global routing, including initial routing and rip-up and reroute, are presented. In Section VI, the experimental evaluation is discussed. Finally, in Section VII, the main conclusion is drawn.

\section{Background}
\label{sec:rework}

\subsection{Related Work}
\label{sec:rwork}

\textbf{OARSMT Generation:} The OARSMT problem has been researched for many years. In the last century, the work~\cite{Clarkson87} proposed the approach to calculate the shortest connection between two pins in a layout with obstacles. Recently, with the increasing constraints in new technologies, OARSMT has become more important. For example, work~\cite{yang03} first constructs an RSMT without considering obstacles, and then uses a four-step algorithm to legalize the edges that intersect with the obstacles. However, the scenarios of the legalization are relatively straightforward and do not globally consider enough conditions. Work~\cite{Lin08} proposes a solution using the obstacle-avoiding spanning graph (OASG), which trims edges and nodes to yield a minimum spanning tree for an acceptable quality OARSMT in polynomial time. However, it may overlook optimal solutions early on, resulting in excessive wirelength, and has impractical runtime for complex cases. In contrast, work~\cite{Ajwani11} also utilizes OASG but incorporates FLUTE for fine-tuning, achieving better quality and runtime, but misses the early global guidance from FLUTE. Work~\cite{Feng06} approaches the OARSMT problem using the $\lambda$-geometry plane. Subsequent works~\cite{Lin08_2, Ghosal13, Ghosal13_2,Liu14,Lin18} address multi-layer OARSMT, with work~\cite{Lin18} being high quality by leveraging intermediate information of maze routing to optimize tree costs, reducing runtime while improving quality. Additionally, work~\cite{Chen22} explores machine learning, employing a reinforcement learning algorithm to generate tree structures for small cases (12 pins, 20 obstacles) within limited runtimes. Compared with the previous works, leveraging the predefined rules, the proposed OARSMT algorithm has overall better quality, and has limited runtime, especially in large cases. 

\textbf{Global Routing:} Global routing is one of the critical steps in physical design, and there are many studies investigating efficient global routing. For example, FastRoute 1.0~\cite{fasttoute1} constructs a Steiner tree based on congestion cost, further reducing congestion through edge-shift techniques, and ultimately achieves a high-quality solution by maze routing algorithms. FastRoute 4.0~\cite{fasttoute4} leverages the via aware Steiner tree generation, 3-bend routing and layer assignment to reduce the via count and runtime. NTHU-Route~\cite{nthu1} utilizes iterative rip-up and reroute with innovative techniques to improve routing efficiency. To enhance solution quality and runtime performance, NTHU-Route 2.0~\cite{nthu2} incorporates a new history-based cost function, new ordering methods for congested region identification and two implementation techniques. FGR~\cite{fgr} utilizes 3D maze routing based on discrete Lagrange multipliers to reroute net for an existing routing solution, GRIP~\cite{grip} leverages integer linear programming to choose an optimal route for each net. These two methods run too slowly to be practical. MGR~\cite{mgr} resorts to an efficient multi-level framework to reroute nets in the congested region on the 3D grid graph with comparable runtime of 2D. Recently, CUGR 2.0~\cite{cugr2} uses DAG-based 3D pattern routing and sparse grid graph maze routing. Compared to majority routing algorithms, CUGR 2.0 not only demonstrates a substantial improvement in the quality of results but also significantly reduces the running time, showcasing enhanced computational efficiency. However, the above mentioned methods have not specifically considered the wide presence of obstacles with arbitrary numbers and sizes, leading to inefficiency of solving the violations by the obstacles.

To address the aforementioned issues, we propose a novel obstacle-avoiding global routing flow. This flow involves using our proposed fast and efficient OARSMT algorithm during the initial routing phase, and employing OARSMT-guided sparse maze routing and obstacle-aware sparse maze routing during the rip-up and reroute phase. Experimental results demonstrate that this flow is effective when obstacles are widely involved.

\subsection{Problem Formulation}
\label{sec:formulation}

In 3D global routing, the layout area is typically partitioned into an array of rectangular cells called global routing cells (GCells), which is represented as a 3D grid graph $G(V,E)$. Each GCell is treated as a vertex ($v\enspace \epsilon \enspace V$). A complete input case $C$ consists of lots of nets, with each net containing $m$ pins $\{p_0, p_1, p_2, ..., p_{m-1}\}$. Additionally, there are $n$ obstacles $\{b_0, b_1, b_2, ..., b_{n-1}\}$ that are all rectangles with width $w_i$ and height $h_i$ for each $b_i$. The OARSMT problem and the global routing problem based on the obstacle-avoiding feature are defined as follows.

Following the typical previous works~\cite{Lin08,Lin18,Zhang23}, the OARSMT problem used in global routing tree generation is to find a rectilinear Steiner tree with minimal wirelength that connects all pins, with no edge crossing any obstacles and no node located inside any obstacles. Note that edges passing through the boundaries are legal. The pins and obstacles are located on a $xy$-plane without overlap. Therefore, when executing OARSMT, it is necessary to map the obstacles in 3D space to 2D space. The cost of the tree is the wirelength, and an edge passing through (or a node located at) the boundary of an obstacle is assumed to be legal.

For the complete global routing problem with obstacles, the capacity of an edge between two adjacent GCells represents the number of available routing tracks through it. The demand of an edge represents the capacity that has already been used on that specific edge. If the demand of an edge exceeds its capacity, it results in an overflow. It is established that obstacles are not traversable by route of net. As a consequence, the capacity of the area where obstacles are positioned is set to zero. The primary objective of obstacle-avoiding global routing is to minimize the $cost(C)=WL + \alpha_{ow} OW + \alpha_{ov} OV$, where $WL$ is the total wirelength of all nets, $OW$ is the total overflow cost, and $OV$ is the cost of obstacle violations. Note that since an obstacle is not traversable, its violation weight $\alpha_{ov}$ is sufficiently high.

For a better description, we use $(x,y)$ to represent a coordinate. Meanwhile, $[n_i,n_j]$ represents a straight line between node $i$ and $j$. $n_i.x$ and $n_i.y$ stand for the x and y coordinates of node $i$. Moreover, $[n_i,dir)$ represents a ray segment line from node $n_i$ with direction $dir$ (A ray segment is a straight line extending from a point indefinitely in one direction only).

\section{Proposed Global Routing Flow Overview}
\label{sec:overview}

\begin{figure}[!hbt]
	\centering
	\includegraphics[width=0.8\columnwidth]{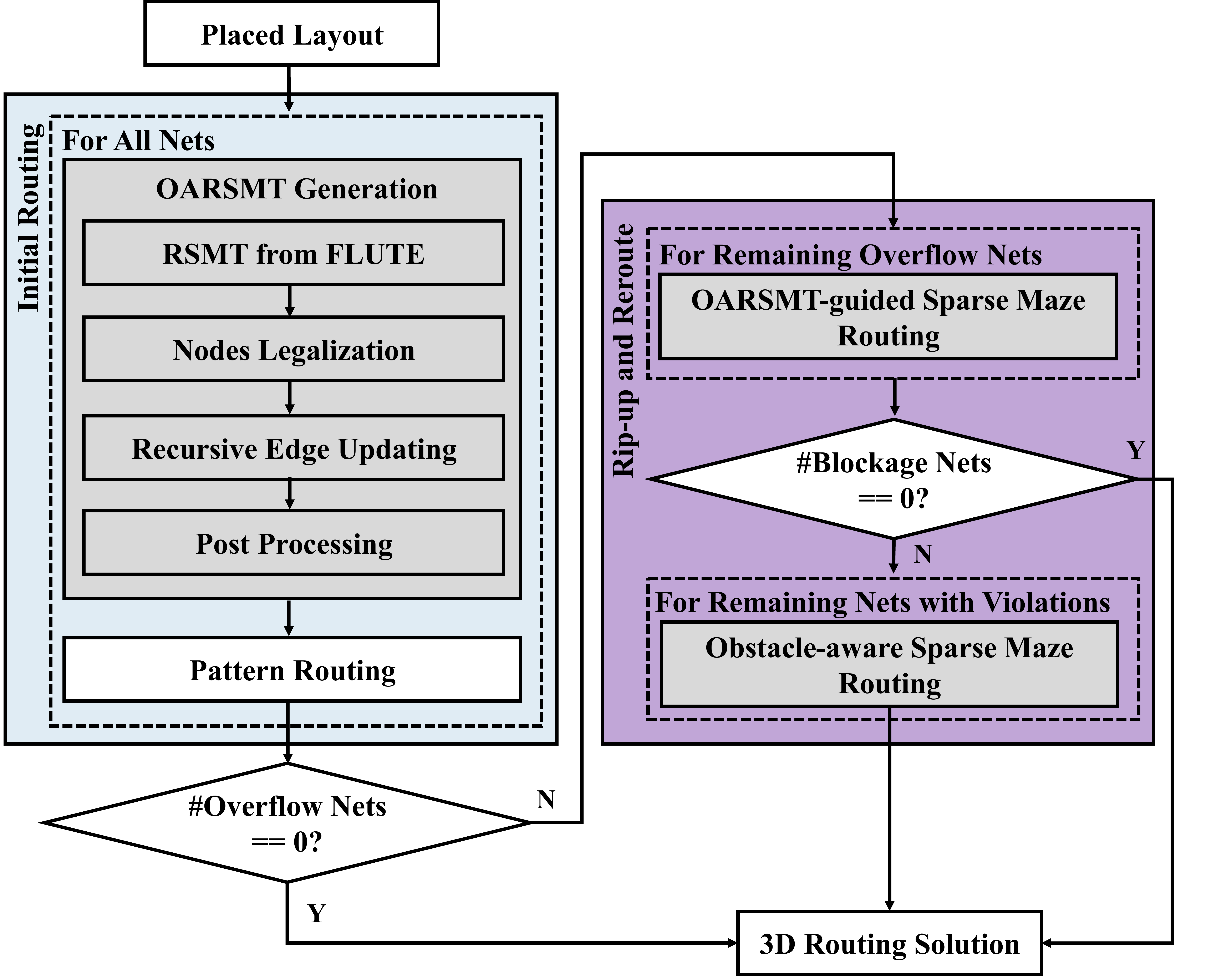}
	\caption{The proposed routing flow. (The proposed algorithms are shown as gray rectangles.)}
	\label{fig:routing_flow}
        
	\vspace{-2ex}
\end{figure}

The proposed global routing flow avoids obstacles during routing with obstacle-aware tree generation and two stages of maze routing. These steps are illustrated in Figure~\ref{fig:routing_flow}. The key steps highlighted in gray are explained in more detail below.

\textit{(1) OARSMT Generation:} To minimize the number of net connections obstructed by obstacles during the initial routing, we construct an OARSMT in the tree generation stage. The OARSMT algorithm begins by using the FLUTE algorithm to determine the structure of the Rectilinear Steiner Minimum Tree (RSMT). Then, the algorithm updates the nodes (steiner nodes and L-shape corner nodes) and edges that intersect with obstacles using a very fast newly proposed rule-based method, and completes the OARSMT generation. Then, an initial route of the net is obtained through pattern routing. Compared to traditional initial routing methods, the proposed approach demonstrates a significant reduction in the number of nets that encounter obstacles, all while maintaining high processing speeds. This improvement can enhance the subsequent obstacle-avoiding processes and offer valuable guidance for the later stages of routing.

\textit{(2) OARSMT-Guided Sparse Maze Routing:} Although the number of nets with obstacle violations has significantly improved after initial routing, there are still many overflow issues and some violated nets. We propose a method called OARSMT-guided sparse maze routing. This method utilizes the OARSMT structure generated during the initial routing as a guide. It incorporates relevant paths into a sparse graph and performs maze routing within this graph. Compared to traditional sparse maze routing, our approach has a higher likelihood of avoiding obstacles, effectively reduces overflow, and achieves a speed comparable to that of standard sparse maze routing.

\textit{(3) Obstacle-aware Sparse Maze Routing:} If there are still nets that have not successfully avoided obstacles, we will perform more refined maze routing for these nets, referred to as obstacle-aware sparse maze routing in this study. The obstacle-aware sparse grid graph will be constructed first, incorporating the boundaries of all obstacles, and maze routing will be performed within this graph. After this phase, the vast majority of nets are able to successfully avoid obstacles.

\section{OARSMT Algorithm Design}
\label{sec:design}

The first key process is the OARSMT generation during the initial routing. With a large number of nets, the proposed OARSMT algorithm can quickly generate tree structures that avoid obstacles. The generated OARSMTs can reduce the obstacle violations at the early step, which can release the workload of later steps and provide guidance for maze routing to achieve better quality and runtime. In the following subsections, the design rationale of the proposed OARSMT algorithm is introduced, followed by a detailed elaboration of the algorithm and flow.

\subsection{Design Rationale and Edge Updating}
\label{sec:edgeupdate}

As FLUTE does not consider obstacles, previous OARSMT works do not apply it in the early stage. This may leave out the valuable guidance of optimized tree structures, and incur more runtime. The proposed algorithm in this work leverages the solutions of FLUTE as guidance, and further updates them to optimize and legalize the layout with obstacles. 

\begin{figure}[!hbt]
        \vspace{-2ex}
	\centering
	\includegraphics[width=0.85\columnwidth]{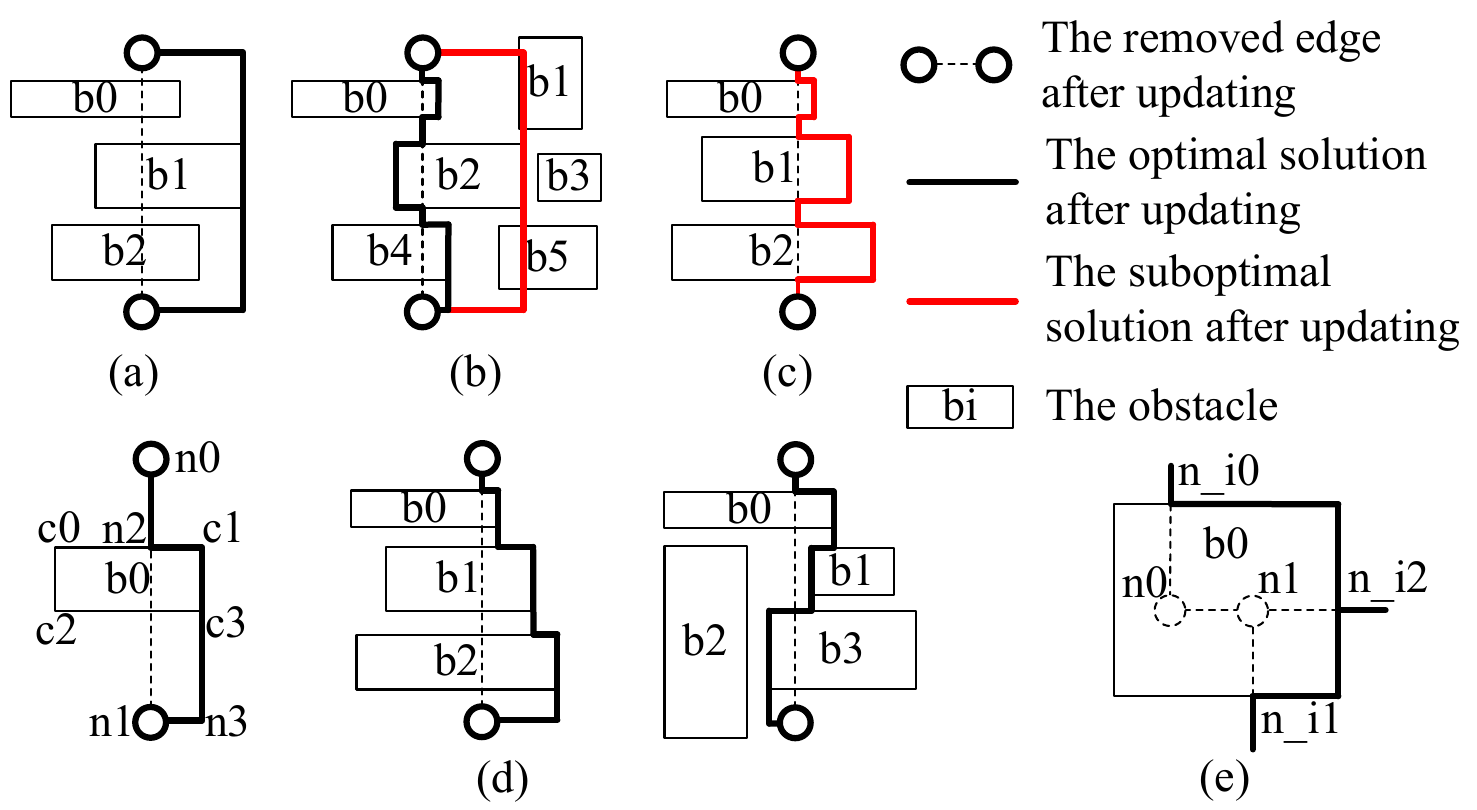}
	\caption{(a) An example of directly detouring; (b) An example where directly detouring is invalid, while connecting through obstacle boundaries is optimal; (c) An example where connecting through obstacle boundaries incurs large cost; (d) The examples where the proposed reference line-based edge updating show the optimal solutions; (e) An example where both the Steiner node and L-shape corner node are located inside an obstacle.}
	\label{fig:ruleex}
	\vspace{-2ex}
\end{figure}

The basic procedure to perform the solution updating is to find all the straight-line edges based on a FLUTE solution (with L-shapes in both directions  considered for two nodes that a single horizontal or vertical line cannot connect) that cross obstacles, and update them one by one. 
Using maze routing or recording the patterns of obstacles in a look-up table might be two approaches, but they either suffer from long runtime or tremendous patterns to be recorded. A trade-off approach can be the rule-based edge updating. Examples about two straightforward rules to update the edges are shown in Figure~\ref{fig:ruleex}(a)-(c), including directly detouring around the obstacles, and connecting through the obstacle boundaries. However, directly detouring is too aggressive, while connecting through obstacle boundaries is too conservative. They can incur violations or large cost, indicated as red lines in Figure~\ref{fig:ruleex}(b)-(c).

\begin{algorithm}[t]
\caption{Edge updating algorithm} 
\raggedright
{\bf Function:}  
Edge\_Update\\
{\bf Input:}  
Source, Target Nodes $n_s$, $n_t$; Reference Line $l_r$; Obstacles $S_b$.\\ 
{\bf Output:}  
Updated Edges $U_e$.
\begin{algorithmic}[1]
\State $U_e \gets \emptyset$
\State $n^\to \gets n_s$
\State $rdir \gets$ $[n_s,n_t]$'s direction
\While{True}
    \State $b_1 \gets$ The 1st obstacle $\in S_b$ that blocks $[n^\to,rdir)$
    \If{$b_1 != NULL$}
        \State $[c_{n0}, c_{n1}]\gets $ $b_1$'s 1st boundary that blocks $[n^\to,rdir)$
        \State $n^\dashv \gets $($rdir$ is horizontal)$ ? (c_{n0}.x,n^\to.y):(n^\to.x, c_{n0}.y)$
        \State $U_e \gets U_e \bigcup \{[n^\to,n^\dashv]\}$
        \State $n^\to \gets n^\dashv$
        \State $n^\dashv \gets dis(c_{n0},l_r)<dis(c_{n1},l_r) ? c_{n0} : c_{n1}$
        \State $U_e \gets U_e \bigcup \{[n^\to,n^\dashv]\}$
        \State $n^\to \gets n^\dashv$
    \Else
        \State $n^\dashv \gets $($rdir$ is horizontal)$ ? (n_t.x,n^\to.y):(n^\to.x, n_t.y)$
        \State $U_e \gets U_e \bigcup \{[n^\to,n^\dashv]\}$
        \State $U_e \gets U_e \bigcup \{[n^\dashv,n_t]\}$
        \State Break
    \EndIf
\EndWhile

\State Return $U_e$
\end{algorithmic}
\label{alg:edge}
\end{algorithm}

To tackle the above challenges while keeping the fast process, a new rule-based edge updating approach is proposed in Algorithm~\ref{alg:edge}, which keeps a balance between the two rules in Figure~\ref{fig:ruleex}(a)-(c). 

We first show an example in Figure~\ref{fig:ruleex}(d). 
To update an edge $[n_0,n_1]$ that crosses obstacles, the proposed algorithm extends the edge line from $n_0$ to $n_1$, and iteratively bends the line when it is blocked on its extending direction. The edge always bends toward the obstacle corner nearest to the original edge.
Taking the left part of Figure~\ref{fig:ruleex}(d) as an example, the line of the original edge $[n_0,n_1]$ based on FLUTE is named the \textit{reference line}. The nodes of the original edge $[n_0,n_1]$ are named source node $n_0$ and target node $n_1$, respectively.
The direction from the source node to the target node is named the \textit{reference direction}, which is vertical down in this example.
A line is initially created and extended from $n_0$ to $n_1$ along the reference direction. Once the extended line ($[n_0,n_2]$) is blocked by an obstacle's boundary ($[c_0,c_1]$ of $b_0$), it is bent and extended to the boundary's corner ($c_1$) that is closer to the reference line, resulting in a line segment $[n_2,c_1]$. Then, it is bent again to extend through the reference direction, resulting in $[c_1,n_3]$. Once the line is extended to reach the same coordinate on the reference direction as the target node $n_1$ (e.g., $n_3$ has the same y-coordinate as $n_1$), the line is bent again towards $n_1$ and connects it.
This edge from the final bending (such as $[n_3,n_1]$) might still cross a few obstacles, which can be solved by iteratively performing the edge updating.

The detailed algorithm is shown as Algorithm~\ref{alg:edge}, which can update edge $[n_s, n_t]$ according to the reference line $l_r$, given obstacles $S_b$, and output the updated edges $U_e$ (For better description, the information of the nodes is also assumed included in the information of edges). 
During initialization, $U_e$ is set to be empty. A node $n^\to$ is maintained to be the current node to be extended from. It is initialized as $n_s$ at Line 2 of Algorithm~\ref{alg:edge}. The reference direction $rdir$ is initialized at Line 3. 
Next, similar to the example in Figure~\ref{fig:ruleex}(d), the edge is extended through $rdir$ like a ray segment $[n^\to,rdir)$ until blocked by an obstacle $b_1$ (Line 5). Then, the first boundary of $b_1$ that blocks $[n^\to,rdir)$ is extracted (Line 7). For example, in Figure~\ref{fig:ruleex}(d), when $n^\to$ is $n_0$, $[c_{n0}$,$c_{n1}]$ is $[c_0$,$c_1]$. Then, $n^\dashv$ is the node where the current edge extension is end. If the reference direction is horizontal, $n^\dashv$'s coordinate contains $c_{n0}$'s x-coordinate and $n^\to$'s y-coordinate. Otherwise, $n^\dashv$'s coordinate contains $n^\to$'s x-coordinate and $c_{n0}$'s y-coordinate (Line 8). Now, current edge extension result $[n^\to,n^\dashv]$ is recorded into $U_e$ (Line 9). The next part is to bend the edge extension to a $b_1$'s corner that is closer to $l_r$, and record the bent line segment as a new edge (Lines 10-12). Finally, $n^\to$ is updated to the reached corner and the procedure is performed again (Line 13). 
When the edge extension from $n^\to$ is never blocked by further obstacles, the edge is extended to the $n^\dashv$ node with the same coordinate in the reference direction as $n_t$ (Line 15). The current edge is then recorded (Line 16), and the edge from the $n^\dashv$ to the target node is finally recorded to finish the connection (Line 17). The edge updating is finished (Line 18).

In addition to the edge itself, both the Steiner nodes and the L-shape corner nodes along the edges also need to be updated when they are located inside the obstacles, such as the example in Figure~\ref{fig:ruleex}(e). This can be fixed by regenerating the edges shown as the black lines in the figure, and removing the invalid edges and nodes. 
The black lines are generated as the shortest edges placed around the obstacle that can connect all the intersection points ($n_{i0}$-$n_{i2}$).

\subsection{Rule Enhancements}

\label{sec:extend}

\begin{figure}[!hbt]
	\centering
	\includegraphics[width=0.8\columnwidth]{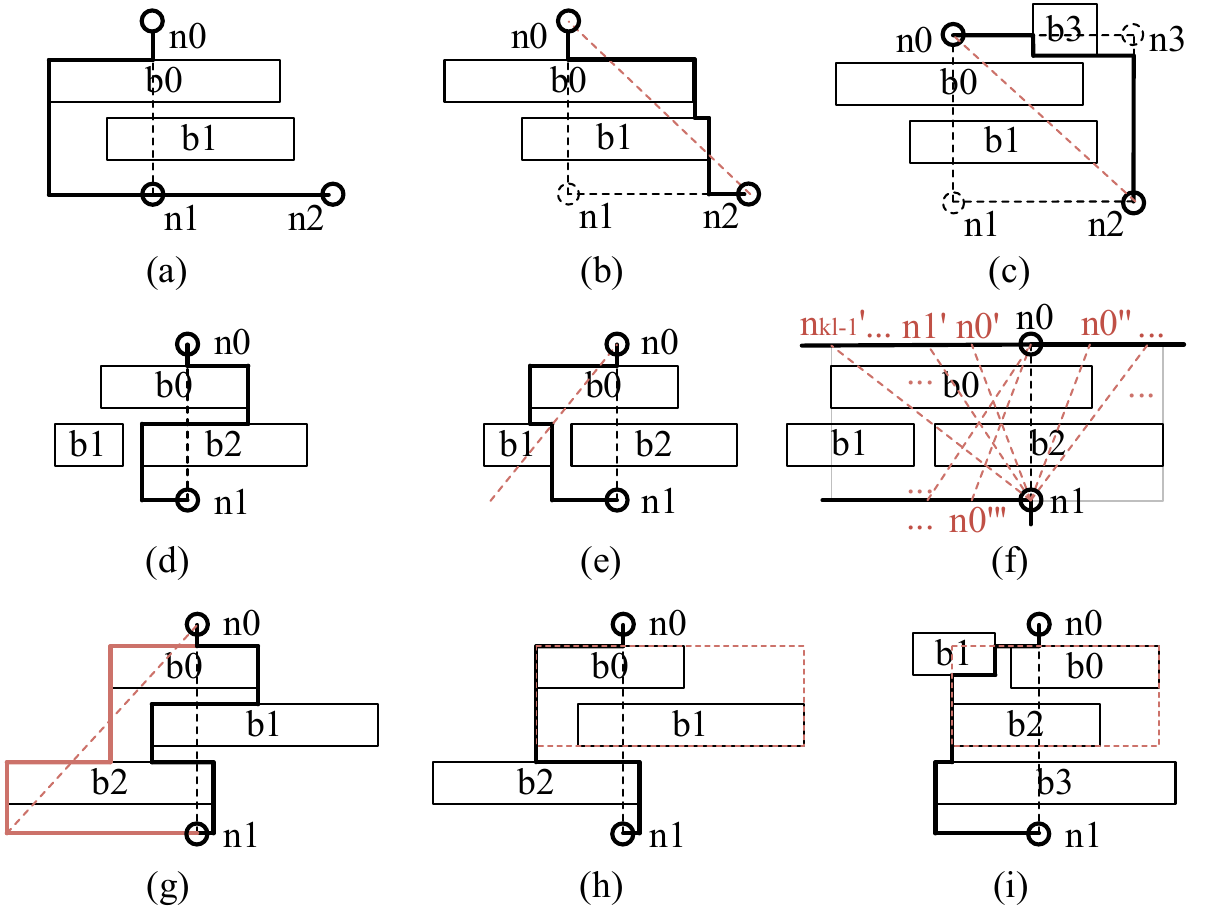}
	\caption{(a)-(c) An example of better-updating L-shape connections. (d)-(f) An example of the benefit by sloping the reference line. (g)-(i) An example of the benefit by merging obstacles. (The reference line is shown as the red dotted line if it is not overlapped with the original edge to be updated, and the red solid lines are the its corresponding updated edges, if illustrated.)}
	\label{fig:enhance}        
	\vspace{-2ex}
\end{figure}

The enhanced rules are derived from Algorithm~\ref{alg:edge} Edge\_Update($n_s, n_t, l_r, S_b$), with modifications to the parameters $n_s$, $n_t$, $l_r$, and $S_b$.

\textit{ER1 L-Shape Optimization: }Although edge updating performs well for the cases in Figure~\ref{fig:ruleex}, there are still cases where basic edge updating alone is insufficient. For example, in Figure~\ref{fig:enhance}(a), an L-shape connection generated based on FLUTE is shown. Here, $n_0$ and $n_2$ are two tree nodes connected by an L-shaped tree edge, while $n_1$ represents a corner node for this connection. Edge updating needs to update $[n_0,n_1]$, and the solution is Figure~\ref{fig:enhance}(a) as the left corner of $b_0$ is closer to $[n_0,n_1]$. However, if $n_0$ connects $n_2$ by the edges on the right of Figure~\ref{fig:enhance}(b), after $[n_0,n_1]$ and a segment of $[n_1,n_2]$ are removed, the optimal solution can be obtained. This can be realized by setting the reference line as $[n_0,n_2]$. Moreover, changing the L-shape direction (where $n_3$ is also a corner node) as shown in Figure~\ref{fig:enhance}(c) might also yield another optimal solution, even when there is a new obstacle $b_3$. Inspired by these findings, given an edge $[n_i,n_j]$ needs to be updated, if the degree of $n_i$ or $n_j$ is 2 and the other incident edge is orthogonal to $[n_i,n_j]$ (which means there exists an L-shape), the reference line of edge updating is the diagonal of this L-shape. Besides, two types of L-shapes are both tried to perform the edge updating, and the solution with a smaller wirelength is kept.

\textit{ER2 Reference Line Sloping: }Besides for L-shapes, there are also other cases that can be explored. For example, Figure~\ref{fig:enhance}(d) shows the solution of the proposed edge updating for $[n_0,n_1]$, but a better solution is shown in Figure~\ref{fig:enhance}(e). This can be obtained by setting the reference line with a slope to the left. In this case, given a general edge $[n_i, n_j]$ (such as $[n_0, n_1]$ in Figure~\ref{fig:enhance}(f)) that needs to be updated, the orthogonal edges to $[n_i, n_j]$ from node $n_i$ or $n_j$ are checked. Note that if no orthogonal edge exists in a certain direction of $n_i$ (or $n_j$), the sloping reference line from $n_j$ (or $n_i$) in that direction is not considered, as this may introduce additional edges and lead to higher cost. Such as two black edges from $n_0$ to its left and right, and one edge from $n_1$ to its left in Figure~\ref{fig:enhance}(f). 
The ray segments covering the checked edges from $n_i$ or $n_j$ are obtained, such as $[n_0, left)$, $[n_0, right)$, and $[n_1, left)$.
For each obtained ray segment, $k_l$ nodes in total are used for reference line building, such as $n_0'$ to $n_{k-1}'$. These $k_l$ nodes are named \textit{hook nodes}.
Next, multiple reference lines $l_r$ for edge updating are built from $n_i$ (or $n_j$) to the hook nodes on the obtained ray segments from $n_j$ (or $n_i$), respectively. Such as reference lines $[n_1,n_0']$, $[n_1,n_1']$, etc.
Note that the spacing between each nearest pair of hook nodes on the same ray segment is the same. Besides, all hook nodes are within the bounding box of $[n_i, n_j]$ (the minimal rectangular region that encloses all obstacles overlapping with $[n_i, n_j]$, such as the gray rectangle in Figure~\ref{fig:enhance}(f)).
Finally, each reference line $l_r$ built from $n_i$ (or $n_j$) to a hook node is separately tried for Edge\_Update($n_s$,$n_t$,$l_r$,$S_b$) with $n_s\gets n_i$ (or $n_j$) and $n_t\gets n_j$ (or $n_i$), respectively. The solution with the smallest local wirelength is kept. 

\textit{ER3 Obstacle Merging: } 
Another approach for rule enhancement is to merge multiple nearby obstacles, and the bounding box of these merged obstacles is regarded as a new obstacle when applying edge updating. In Figure~\ref{fig:enhance}(g), the original edge updating incurs large cost due to lots of bends, and the sloped reference line on the left can also incur large cost.
In contrast, the solution in Figure~\ref{fig:enhance}(h) becomes more optimized where $b_0$ and $b_1$ are merged. Note that sometimes merging obstacles can cause overlaps, such as $b_1$ in Figure~\ref{fig:enhance}(i). For efficiency, if the edge updating is blocked by the overlapped obstacles, the edge extends on the boundaries of the overlapped obstacles, through the shortest path until able to perform the edge extension without the overlaps. Finally, totally $k_m+1$ ways of merging are tried for each edge $[n_s,n_t]$ to be updated as follows, where $k_m$ is user defined parameter.
Assume there are $n'$ successive obstacles $b_0,b_1,...b_{n'-1}$ crossing edge $[n_s,n_t]$, then each group of $n_m$ successive obstacles can be temporarily merged. For example, if $n_m=3$, the obstacles in each group, such as $\{b_0,b_1,b_2\}$, $\{b_3,b_4,b_5\}$, $\{b_6,b_7,b_8\}$, etc., are separately merged as one obstacle. 
The value $n_m$ is tried multiple times as $1$, $n'/k_m$, $2n'/k_m$, $k_m\times n'/k_m$. For each $n_m$ value tried, a temporary new obstacle set $S_b$ is obtained, and the edge updating for $[n_s,n_t]$ by Edge\_Update($n_s$,$n_t$,$l_r$,$S_b$) is performed.
The solution with the smallest local wirelength among $k_m+1$ times trying is kept.

When updating an edge, the 3 enhanced rules are combined to be tried and the solution with the smallest cost (wirelength) is kept. This combination will be introduced in Algorithm~\ref{alg:oarsmt} later.

\subsection{The Rule-Based OARSMT Algorithm}
\label{sec:rulebasedoarsmt}

\begin{figure}[!hbt]
\vspace{-2ex}
	\centering
	\includegraphics[width=0.7\columnwidth]{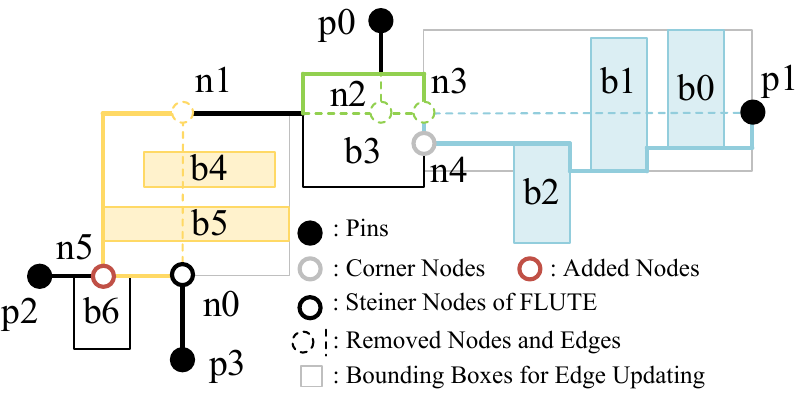}
	\caption{An example of the solution from the proposed algorithm. (Different colors stand for different elements considered in different processes)}
	\label{fig:mainex}
        
\end{figure}

Based on the edge updating rule, the proposed rule-based OARSMT algorithm is proposed in Algorithm~\ref{alg:oarsmt}, where the inputs include a set of pins $S_p$ and a set of obstacles $S_b$. It outputs a set of edges $S_e$ as the solution. The flow can be separated into 4 steps.

\begin{algorithm}[t]
\caption{Rule-based OARSMT algorithm} 
\raggedright
{\bf Function:}  
OARSMT\_Generation\\
{\bf Input:}  
Pins $S_p$; Obstacles $S_b$.\\ 
{\bf Output:}  
Edges that connect all pins $S_e$.
\begin{algorithmic}[1]

\State $S_{Fe} \gets$ Run FLUTE to get the solution of edges
\State $S_e \gets$ Build the rectilinear tree based on $S_{Fe}$
\For{Each obstacle $b$ in $S_b$}
    \State $S_n \gets$ The set of nodes locating inside $b$
    \State Revise edges in $S_e$ related to $S_n$.
\EndFor
\For{Each edge $e$ in $S_e$}
    \State $S_{be} \gets$ The set of obstacles that blocks $e$
    \If{$S_{be}== \emptyset$}
        \State Continue
    \EndIf
    \State $B_b \gets$ The bounding box of all obstacles in $S_{be}$
    \State $S_{candb} \gets$ The set of obstacles that overlap $B_b$
    \State $U_{emin} \gets \emptyset$, $WL(U_{emin})\gets \infty$
    \If{$e$ is on an L-shape pattern $L$} //\textbf{ER1}
        \State $l_r \gets$ the diagonal of $L$
        \State $n_s,n_t\gets$ 2 end nodes of $e$
        \State $U_e\gets$ Edge\_Update($n_s,n_t,l_r,S_{candb}$)
        \State $L'\gets$ the L-shape by flipping $L$ to another direction
        \State $e'\gets$ the edge on $L'$, with a same end node as $e$
        \State Calculate $U_{e'}$ by performing Lines 8, 12-13, 16-18 by replacing $e$ with $e'$
        \State \textbf{if} $WL(U_e)<WL(U'_e)$, $U_{emin}\gets U_e$ \textbf{else} ($U_{emin}\gets U'_e$ and $e\gets e'$)
    \Else
    \For{Each output $S_{mb}$ from ER3 when merging $S_{candb}$} //\textbf{ER3}
        \For{Each possible $l_r$ and the source-target node pair ($n_s,n_t$) of $e$ by ER2} //\textbf{ER2}
            \State $U_e \gets$ Edge\_Update($n_s,n_t,l_r,S_{mb}$)
            \State \textbf{if} $WL(U_e)<WL(U_{emin})$, $U_{emin}\gets U_e$ 
        \EndFor
    \EndFor
    \EndIf
    \State $S_e \gets S_e\setminus e$
    \State $S_e \gets S_e\bigcup U_{emin}$
\EndFor

\State Remove redundant edges, merge overlapped edges in $S_e$

\State Return $S_e$
\end{algorithmic}
\label{alg:oarsmt}
\end{algorithm}

\textit{(1) RSMT Generation: }For the first step, the FLUTE solution is obtained (Line 1 of Algorithm~\ref{alg:oarsmt}). This solution usually only includes the connections between nodes, and thus a rectilinear tree first needs to be built as $S_e$ in Line 2. Each edge between two connected nodes from FLUTE is constructed by an L-shape pattern, and the direction is basically randomly selected at first.

\textit{(2) Nodes Legalization: }The second step is to legalize the nodes (Steiner nodes and L-shape corner nodes) inside the obstacles (Lines 3-6) as Figure~\ref{fig:ruleex}(e). 
For a set of invalid nodes $S_n$ within an obstacle $b$, the intersection points between $b$'s boundaries and edges connecting nodes in $S_n$ are found. Then, the shortest edges placed around the obstacle boundaries that are able to connect all such intersections are calculated and added to edge set $S_e$. Finally, the edge segments inside $b$ are removed from $S_e$.
An example is also shown as the green parts in Figure~\ref{fig:mainex}.

\textit{(3) Iterative Edge Updating: }The critical step is edge updating, which is iteratively performed (Lines 7-33). Each time an edge $e$ with violations is selected, the set of obstacles $S_{be}$ that block $e$ is obtained (Line 8). Then, a bounding box $B_b$ is heuristically introduced to greatly reduce the complexity (Line 12). The bounding box is a rectangle that just surrounds $e$ and all obstacles in $S_{be}$, and the candidate obstacles $S_{candb}$ are selected as the obstacles overlap $B_b$ (Line 13). For example, in Figure~\ref{fig:mainex}, when updating edge $[n_0,n_1]$, $B_b$ is built based on $[n_0,n_1]$, $b_4$ and $b_5$. In this case, $S_{candb}=\{b_4,b_5\}$. In contrast, when updating edge $[p_1,n_3]$, although $B_b$ is not built based on $b_2$, but $b_2$ is included in $S_{candb}$ since $B_b$ overlap $b_2$.
Next, a record of the minimum wirelength connection is initiated at Line 14.
The enhanced rule ER1 is first checked at Line 15. If $e$ is on an L-shape pattern, the reference line $l_r$, the source node $n_s$, and the target node $n_t$ are extracted as Lines 16-17. The edge updating from Algorithm~\ref{alg:edge} is called to obtain the updated edges $U_e$. Then, according to ER1, the edge of another direction of L-shape is tried to be updated (Lines 19-21), and the better solution is kept (Line 22).
Otherwise, if $e$ is not on an L-shape, the edge updating is conducted (Line 26) by trying all combinations of $n_s$, $n_t$, $l_r$, and $S_{mb}$ from ER2 and ER3 (Lines 24-25)

Once a solution with better wirelength is found, the record $U_{emin}$ is updated (Line 27). The best solution is selected to replace edge $e$ (Lines 31-32).
Note that all updated new edges will be further checked after edge updating of $e$, as they might also cross obstacles, so finally all edges are legal.

\textit{(4) Post Processing: }Finally, each edge connecting a node (excluding pins) with degree 1 is removed, and the overlapped edges are merged (Line 34). This can be finished quickly by checking the edges connected to each node (a Steiner node, an L-shape corner node, or a pin). Additionally, we perform cycle detection on the generated tree. Even if a non-tree structure is produced, it can be quickly repaired using classical algorithms. In our experiments, the probability of generating a non-tree structure is $\sim$1\%.

\subsection{Data Structure Improvement and Time Complexity}
\label{sec:datastru}

\begin{figure}[!hbt]
	\centering
	\includegraphics[width=0.7\columnwidth]{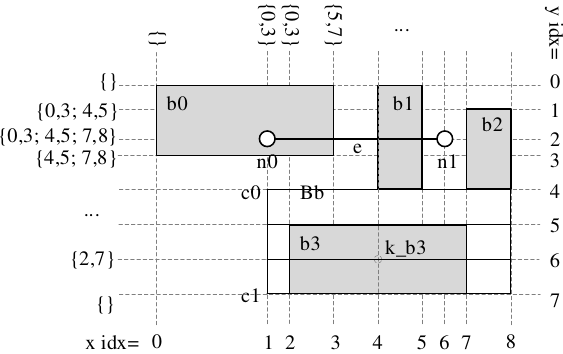}
	\caption{An example of the improved data structure. (The grids based on the centers of obstacles are not completely shown in this example for more clearly illustrating. Only the center $k_{b3}$ of $b_3$ is shown.)}
	\label{fig:data}
        
	\vspace{-2ex}
\end{figure}

The critical and the most time-consuming part is the iterative edge updating (Lines 7-23 in Algorithm~\ref{alg:oarsmt}). During this part, frequently checking for if a line crosses any obstacles (and which obstacles) is needed, such as by Line 8 of Algorithm~\ref{alg:oarsmt}, especially when bending an edge during the edge updating in Section~\ref{sec:edgeupdate}. This can be realized by using an R-tree to store obstacles and querying it to check if a segment or rectangle overlaps with any obstacles. However, the R-tree structures are relatively complex and general that also allow the insertions and overlaps of the stored rectangles (obstacles). As obstacles are typically fixed and can be represented by non-overlap rectangles, the R-tree data structure can be replaced by a simpler data structure for higher efficiency. Therefore, we propose a specific data structure to avoid redundant functions and it is illustrated in Figure~\ref{fig:data}. For example, Figure~\ref{fig:data} shows a layout with obstacles indicated as gray rectangles. The Hanan grid can then be generated as the dotted lines. Note that the grid is not only based on the pins and the boundaries of obstacles, but also based on the centers of obstacles such as $k_{b3}$ of $b_3$. For each row and column of the grid, a \textit{range list} is maintained with multiple sorted pairs of numbers, separated by semicolons. Each pair is a range of an obstacle's width or height, respectively. We name each pair as \textit{range pair}. For example, for Row 1 ($y=1$), there are two obstacles $b_0$ and $b_1$, which are located at $0\leq x\leq 3$ and $4\leq x\leq 5$, respectively. In this case, two range pairs are stored as $\{0,3; 4,5\}$. Besides, to track which obstacle each range pair corresponds to, additional lists containing the obstacle pointer for each range pair are maintained, which is not shown in this example.
The range lists are built in $O(n)$ runtime for one time, by checking the locations and ranges of all obstacles and pushing the ranges into the corresponding list.

After the range lists are built, checking which obstacles a given line crosses can be quickly finished. The procedure can be illustrated by checking the line $e$ in Figure~\ref{fig:data}. 
$e$ is horizontally located at Row 2, so the x-coordinates (1 and 6) of $e$'s two nodes $n_0$ and $n_1$ are checked in the range list $\{0,3; 4,5; 7,8\}$ to obtain the list indices where these two x-coordinates are. $n_0$ locates between $x=0$ and $x=3$ (between indices 0 and 1 of $\{0,3; 4,5; 7,8\}$), while $n_1$ locates between $x=5$ and $x=7$ (between indices 3 and 4). 
The obstacles of which the range pairs' indices are between the indices of $n_0$'s and $n_1$'s locations are categorized as crossed the line $e$. In this example, the obstacles cross $e$ are $b_0$ and $b_1$.
The obstacles can be obtained by the maintained obstacle pointers associated with the range pairs. This check only takes $O(log(n))$ time, since each list is sorted and contains at most all $n$ obstacle ranges. 

Furthermore, if all obstacles that a bounding box (such as $B_b$ in Figure~\ref{fig:data}) overlap need to be obtained, a short boundary of the bounding box ($[c_0,c_1]$) can be selected, and each grid line orthogonal to the short boundary inside $B_b$ is then selected, which is shown as the 4 horizontal lines in $B_b$ (including 2 long boundaries and the line through the center of $b_3$). Finally, these selected lines are iteratively checked to get all overlapped obstacles. The runtime is $O(len_{minb}\times log(n))$, where $len_{minb}$ is the length of the short boundary of the bounding box.

Based on this, the overall time complexity is analyzed as follows. The main time-consuming part is edge updating. Since an edge at most crosses $n$ obstacles, the bending of the edge performs at most $n$ times. After each time of bending, we identify the 1st obstacle intersecting the edge’s current extending path (Line 5 of Algorithm~\ref{alg:edge}), 
thus requiring $O(nlog(n))$ runtime in total for each edge updating. When $k_l$ for reference line sloping and $k_m$ for obstacle merging are globally fixed in the tool for any layouts as in our experiments, it does not affect the time complexity.
In Algorithm~\ref{alg:oarsmt}, as there are $m$ pins, the order of the edges in $S_e$ from FLUTE is $O(m)$. This implies that Lines 7–23 are essentially executed $O(m)$ times. 
As lines 8, 12, and 17 needs $O(log(n))$, $O(len_{minb}log(n))$, and $O(nlog(n))$ runtime, the overall time complexity is $O(m\times (log(n)+len_{minb}log(n)+n\times log(n)))$. In practice, usually the maximum size of the support layout for a placer is fixed, so $len_{minb}$ is a constant. In this case, the time complexity is $O(m\times n\times log(n))$. 
This is less than the state-of-the-art previous work~\cite{Lin18}. Besides, in most cases, an edge crosses only a small portion of obstacles, and the $n$ in the time complexity is much smaller than the number of all obstacles. This further leads to fast speed in practice. 

\section{Obstacle-Avoiding Global Routing}
\label{sec:flow}

Based on the proposed OARSMT algorithm, the obstacle-avoiding global routing flow in Section~\ref{sec:overview} is elaborated.

\subsection{Initial Routing}
\label{sec:initial_routing}

\begin{figure}[!hbt]
        \vspace{-2ex}
	\centering
	\includegraphics[width=0.85\columnwidth]{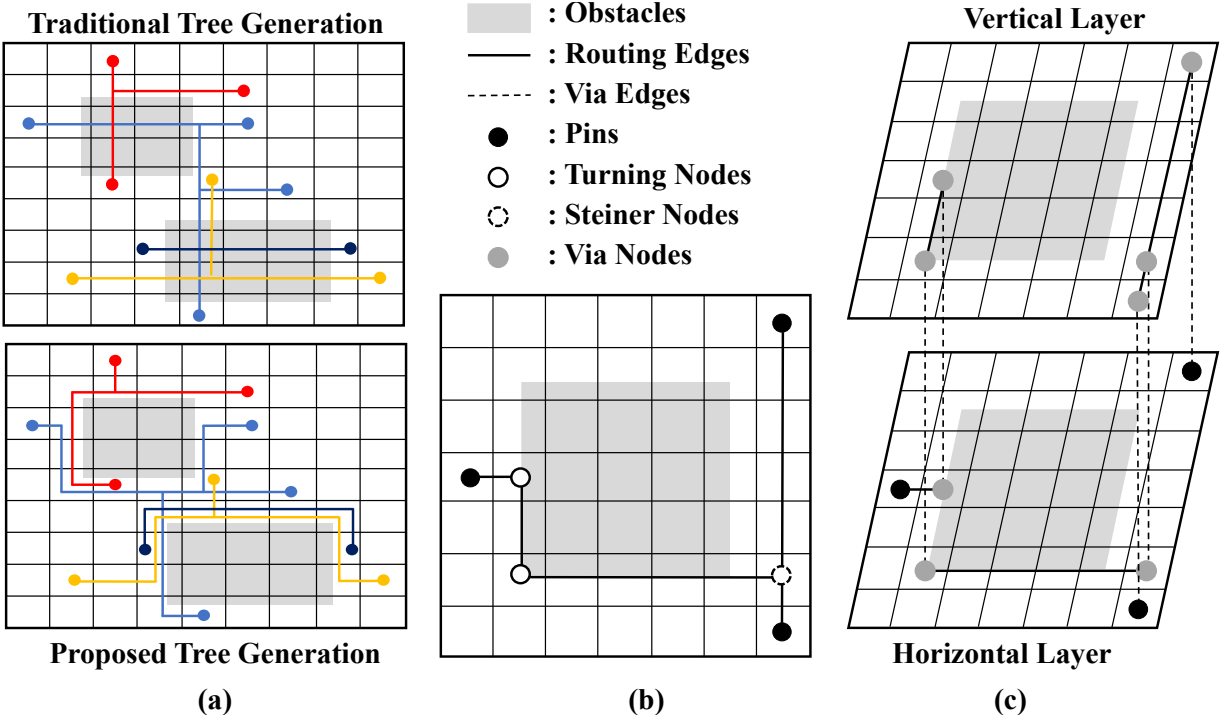}
	\caption{(a) An example of a simple case corresponds to both traditional initial routing and our proposed initial routing. The different colors represent different nets. (b) The corresponding 2D solution of (c). (c) An example of the 3D result obtained under our initial routing.}
	\label{fig:init_routing_ex}
        
	\vspace{-2ex}
\end{figure}

Global routers often overlook obstacles during initial tree generation, leading to violations. The upper diagram of Figure~\ref{fig:init_routing_ex}(a) shows the solutions generated for four multi-pin nets without considering obstacles. This oversight significantly burdens the subsequent phases. In the initial routing stage that we propose, we take into account obstacles during the routing process. By employing the efficient OARSMT algorithm during tree generation, we can greatly reduce the instances of obstacle violations in the initial routing phase while ensuring speed is maintained, as shown in Figure~\ref{fig:init_routing_ex}(a).

The first step of initial routing is to generate OARSMTs for all nets. To reduce runtime while maintaining quality, the OARSMT algorithm only considers obstacles that overlap with the bounding box of each net as part of its inputs. This may result in some nets being in violation, but these issues can be quickly addressed in subsequent steps. After generating the OARSMTs for all nets, traditional pattern routing is applied to determine the routing patterns and the layers of the edges. An example of pattern routing with layer assignment is illustrated in Figure~\ref{fig:init_routing_ex}(b) and (c), which also present the initial routing solution.

Note that although the proposed approach in Figure~\ref{fig:init_routing_ex}(a) places more demands and bends on GCells near obstacles than the traditional one due to avoiding obstacles by OARSMTs, it does not imply that the proposed approach incurs greater congestion, as the traditional approach does not even begin to effectively avoid obstacles at this stage. Actually, the proposed approach has even less congestion in the final results, because most of the obstacles are already well avoided at a very early stage, which alleviates the tasks during rip-up and reroute.

\subsection{Rip-up and Reroute}
\label{sec:rrr_routing}

Although initial routing can help most nets avoid obstacles, further attention is needed to address congestion. Therefore, maze routing is applied to rip up and reroute the remaining overflowed and violated nets. However, maze routing is time-consuming due to the extensive search space involved. One relatively effective method to reduce the search space for maze routing, without significantly compromising the solution quality, is to use sparse graphs. Unfortunately, in many cases, the paths generated on a sparse graph may not provide obstacles-avoiding solutions, as illustrated in Figure~\ref{fig:rg_graph}(a) and (d). To remedy this, we propose OARSMT-guided sparse maze routing and an enhanced obstacle-aware sparse maze routing approach, both of which significantly improve the likelihood of avoiding obstacles while maintaining limited runtime overhead.

\begin{figure}[!hbt]
	\centering
	\includegraphics[width=0.98\columnwidth]{rg_maze_graph4.pdf}
        \caption{Example of OARSMT-guided sparse maze routing and obstacle-aware sparse maze routing.}
	\label{fig:rg_graph}
        
	\vspace{-2ex}
\end{figure}

\textit{(1) OARSMT-Guided Sparse Maze Routing:} In order to significantly reduce overflows while eliminating obstacle violations during the rip-up and reroute phase, we propose OARSMT-guided sparse maze routing in the first stage of rip-up and reroute. During the process of rip-up and reroute for each net, additional sparse grid lines are created by utilizing the OARSMT structure established during the initial routing. These grid lines extend from each edge of the OARSMT and have a specific width. An example is shown with blue lines in Figure~\ref{fig:rg_graph}(b), which follow the extension of the dotted lines. By combining the ordinary sparse graph and the additional grid lines, the OARSMT-guided sparse graph is generated. An example is shown in Figure~\ref{fig:rg_graph}(c). When performing maze routing on the generated graph, a valid solution without any obstacle violations can be obtained, as shown by the solid line example in Figure~\ref{fig:rg_graph}(c). Note that the solution on the OARSMT-guided sparse graph might not be obtainable on an ordinary sparse graph such as Figure~\ref{fig:rg_graph}(a). Besides, each net's additional grid lines are only applied for its own maze routing, so each OARSMT-guided sparse graph has a limited complexity increase compared to the ordinary one. As a result, the proposed method has comparable execution time with better quality than that of the ordinary graph.

\textit{(2) Obstacle-aware Sparse Maze Routing:} After utilizing OARSMT-guided sparse maze routing, the overflow and obstacle-violated nets experience a significant reduction; however, there are still some nets with obstacle violations. As shown in the dotted OARSMT in Figure~\ref{fig:rg_graph}(e), the OARSMT algorithm considers only obstacles within the bounding box during the tree generation stage, which may result in invalid solutions. To fix this problem, the obstacle-aware sparse maze routing can be applied exclusively to the remaining nets with violations. Utilizing the additional grid lines extending from the boundaries of all obstacles and pins for maze routing, by combining these lines with the ordinary sparse graph, an obstacle-aware sparse graph can be created, as shown in Figure~\ref{fig:rg_graph}(e) and (f). The solid lines illustrate a solution for obstacle-aware sparse maze routing. This method ensures that most nets successfully navigate around obstacles. Although the presence of numerous obstacles can slow down the obstacle-aware sparse maze routing process, it is only applied to a limited number of remaining nets. As a result, the overall runtime for this step remains reasonable.

\section{Evaluations}
\label{sec:evaluation}

\subsection{Experiment Setup}

The experiments are conducted on a server with Intel Xeon Gold 6348 CPU. In the experiments of the OARSMT algorithm, there are two types of benchmarks: Randomized testcases and standard benchmarks. Each testcase in benchmarks contains one net to be tested, which follows the regular evaluation process as previous OARSMT works~\cite{Lin18,Lin08,Zhang23}.
For randomized testcase generation, 3 factors are set: The number of pins and obstacles, and the obstacle density on the layout. Other parameters are randomly generated. Each setting is tested 50 times to obtain the average result. Besides, we also test all the standard benchmarks, IND, RC, and RT~\cite{oarsmtcase} involved in the compared works~\cite{Lin18,Lin08,Zhang23}. 
$k_l=5$ and $k_m=2$ in the experiments.
For OARSMT comparison, the previously advanced work proposed by Lin et al.~\cite{Lin18}, a recent work~\cite{Zhang23} and a traditional work~\cite{Lin08} are involved, with works~\cite{Lin18,Lin08} being reproduced.
Similar to recent work~\cite{Chen22}, our experiments also apply the algorithm in work~\cite{Lin18} by 2D mode. The 2D mode of work~\cite{Lin18} does not omit any key techniques from the 3D mode, and thus it still maintains the quality with faster speed.
Besides, as the tree generation in regular global routing flow typically generate 2D trees, applying work~\cite{Lin18} under 2D mode enables the comparison under a complete global routing flow.

In the experiments of obstacle-avoiding global routing, the official open-source implementation of global router CUGR2.0~\cite{cugr2, cugr2_code} is adopted, where the proposed algorithms are implemented to replace the existing routing phases of CUGR2.0. However, our experimental results for wirelength, via count, and overflow cost were obtained by the evaluation function of the official CUGR2.0 implementation without change.
Furthermore, to better demonstrate the effectiveness of our method, obstacle violations are evaluated separately and not included in the overflow cost metric.

ISPD24 contest~\cite{ispd24} benchmark circuits were used in our experiments, as they contain testcases from the regular size to the very large size.
We set the macros in the benchmarks as obstacles for our experiments, of which the areas are forbidden to be crossed by wires at all layers. In addition, we also add 60 random obstacles to each testcase of the benchmarks as additional testcases.

\subsection{OARSMT Comparisons on Randomized Testcases}

\begin{table}[!ht]
\centering
\caption{The comparisons with the previously advanced work~\cite{Lin18} on the randomized testcases.}
\label{tab:comp}

\resizebox{0.85\linewidth}{!}{
\begin{tabular}{|cc|cccccc|}
\hline
\multicolumn{1}{|c|}{\textbf{}}                           & \textbf{Density} & \textbf{\#Obs=10} & \textbf{\#Obs=50} & \textbf{\#Obs=100} & \textbf{\#Obs=500} & \textbf{\#Obs=1000} & \textbf{\#Obs=2000} \\ \hline
\multicolumn{8}{|c|}{\textbf{Speedup}}                                                                                                                                                                     \\ \hline
\multicolumn{1}{|c|}{\multirow{4}{*}{\textbf{\#Pin=10}}}  & \textbf{10\%}    & \textbf{34.56x}   & \textbf{57.03x}   & \textbf{133.74x}   & \textbf{965.18x}   & \textbf{1709.18x}   & \textbf{2711.79x}   \\
\multicolumn{1}{|c|}{}                                    & \textbf{30\%}    & \textbf{28.02x}   & \textbf{42.13x}   & \textbf{65.05x}    & \textbf{436.41x}   & \textbf{1037.00x}   & \textbf{1255.36x}   \\
\multicolumn{1}{|c|}{}                                    & \textbf{50\%}    & \textbf{20.61x}   & \textbf{23.86x}   & \textbf{47.62x}    & \textbf{277.52x}   & \textbf{863.57x}    & \textbf{1052.37x}   \\
\multicolumn{1}{|c|}{}                                    & \textbf{70\%}    & \textbf{19.97x}   & \textbf{18.41x}   & \textbf{31.61x}    & \textbf{261.89x}   & \textbf{949.72x}    & \textbf{725.38x}    \\ \hline
\multicolumn{1}{|c|}{\multirow{4}{*}{\textbf{\#Pin=20}}}  & \textbf{10\%}    & \textbf{50.07x}   & \textbf{99.56x}   & \textbf{143.85x}   & \textbf{1073.18x}  & \textbf{1713.21x}   & \textbf{2332.41x}   \\
\multicolumn{1}{|c|}{}                                    & \textbf{30\%}    & \textbf{37.29x}   & \textbf{63.26x}   & \textbf{85.33x}    & \textbf{402.39x}   & \textbf{793.79x}    & \textbf{1159.35x}   \\
\multicolumn{1}{|c|}{}                                    & \textbf{50\%}    & \textbf{37.37x}   & \textbf{40.62x}   & \textbf{55.80x}    & \textbf{251.26x}   & \textbf{501.56x}    & \textbf{790.10x}    \\
\multicolumn{1}{|c|}{}                                    & \textbf{70\%}    & \textbf{37.41x}   & \textbf{33.13x}   & \textbf{45.35x}    & \textbf{198.97x}   & \textbf{361.32x}    & \textbf{563.91x}    \\ \hline
\multicolumn{1}{|c|}{\multirow{4}{*}{\textbf{\#Pin=30}}}  & \textbf{10\%}    & \textbf{174.33x}  & \textbf{109.19x}  & \textbf{179.57x}   & \textbf{1075.40x}  & \textbf{1718.38x}   & \textbf{2443.06x}   \\
\multicolumn{1}{|c|}{}                                    & \textbf{30\%}    & \textbf{59.79x}   & \textbf{72.72x}   & \textbf{99.10x}    & \textbf{420.68x}   & \textbf{681.07x}    & \textbf{763.08x}    \\
\multicolumn{1}{|c|}{}                                    & \textbf{50\%}    & \textbf{64.34x}   & \textbf{54.72x}   & \textbf{71.49x}    & \textbf{256.15x}   & \textbf{399.53x}    & \textbf{643.97x}    \\
\multicolumn{1}{|c|}{}                                    & \textbf{70\%}    & \textbf{58.02x}   & \textbf{43.52x}   & \textbf{59.27x}    & \textbf{166.03x}   & \textbf{308.12x}    & \textbf{493.75x}    \\ \hline
\multicolumn{1}{|c|}{\multirow{4}{*}{\textbf{\#Pin=50}}}  & \textbf{10\%}    & \textbf{102.39x}  & \textbf{157.40x}  & \textbf{205.02x}   & \textbf{1040.27x}  & \textbf{1647.71x}   & \textbf{1876.00x}   \\
\multicolumn{1}{|c|}{}                                    & \textbf{30\%}    & \textbf{114.99x}  & \textbf{99.48x}   & \textbf{135.85x}   & \textbf{449.44x}   & \textbf{602.45x}    & \textbf{690.97x}    \\
\multicolumn{1}{|c|}{}                                    & \textbf{50\%}    & \textbf{112.97x}  & \textbf{89.93x}   & \textbf{100.86x}   & \textbf{259.32x}   & \textbf{380.30x}    & \textbf{433.12x}    \\
\multicolumn{1}{|c|}{}                                    & \textbf{70\%}    & \textbf{106.00x}  & \textbf{72.53x}   & \textbf{81.37x}    & \textbf{188.36x}   & \textbf{267.95x}    & \textbf{315.92x}    \\ \hline
\multicolumn{1}{|c|}{\multirow{4}{*}{\textbf{\#Pin=100}}} & \textbf{10\%}    & \textbf{206.47x}  & \textbf{230.86x}  & \textbf{282.12x}   & \textbf{1073.62x}  & \textbf{1656.79x}   & \textbf{1916.65x}   \\
\multicolumn{1}{|c|}{}                                    & \textbf{30\%}    & \textbf{212.87x}  & \textbf{185.91x}  & \textbf{214.37x}   & \textbf{492.41x}   & \textbf{668.77x}    & \textbf{619.63x}    \\
\multicolumn{1}{|c|}{}                                    & \textbf{50\%}    & \textbf{214.97x}  & \textbf{168.15x}  & \textbf{190.66x}   & \textbf{338.00x}   & \textbf{436.84x}    & \textbf{441.54x}    \\
\multicolumn{1}{|c|}{}                                    & \textbf{70\%}    & \textbf{206.01x}  & \textbf{151.89x}  & \textbf{157.39x}   & \textbf{252.90x}   & \textbf{287.52x}    & \textbf{261.67x}    \\ \hline
\multicolumn{1}{|c|}{\multirow{4}{*}{\textbf{\#Pin=200}}} & \textbf{10\%}    & \textbf{429.27x}  & \textbf{383.08x}  & \textbf{445.77x}   & \textbf{1042.32x}  & \textbf{1457.32x}   & \textbf{1779.90x}   \\
\multicolumn{1}{|c|}{}                                    & \textbf{30\%}    & \textbf{398.38x}  & \textbf{353.72x}  & \textbf{372.01x}   & \textbf{621.75x}   & \textbf{757.48x}    & \textbf{646.16x}    \\
\multicolumn{1}{|c|}{}                                    & \textbf{50\%}    & \textbf{389.05x}  & \textbf{355.30x}  & \textbf{352.46x}   & \textbf{441.75x}   & \textbf{521.98x}    & \textbf{442.24x}    \\
\multicolumn{1}{|c|}{}                                    & \textbf{70\%}    & \textbf{342.13x}  & \textbf{331.91x}  & \textbf{310.92x}   & \textbf{348.23x}   & \textbf{378.06x}    & \textbf{316.04x}    \\ \hline
\multicolumn{8}{|c|}{\textbf{Wirelength Difference ($<$0 Stands for Improvements)}}                                                                                                                                                    \\ \hline
\multicolumn{1}{|c|}{\multirow{4}{*}{\textbf{\#Pin=10}}}  & \textbf{10\%}    & \textbf{-2.51\%}  & \textbf{-2.42\%}  & \textbf{-2.81\%}   & \textbf{-5.00\%}   & \textbf{-4.26\%}    & \textbf{-5.76\%}    \\
\multicolumn{1}{|c|}{}                                    & \textbf{30\%}    & \textbf{-0.76\%}  & \textbf{-1.68\%}  & \textbf{-2.89\%}   & \textbf{-4.65\%}   & \textbf{-4.44\%}    & \textbf{-4.54\%}    \\
\multicolumn{1}{|c|}{}                                    & \textbf{50\%}    & \textbf{0.56\%}   & \textbf{-0.57\%}  & \textbf{-2.65\%}   & \textbf{-2.83\%}   & \textbf{-3.83\%}    & \textbf{-4.54\%}    \\
\multicolumn{1}{|c|}{}                                    & \textbf{70\%}    & \textbf{-0.31\%}  & \textbf{-4.86\%}  & \textbf{-2.54\%}   & \textbf{-3.53\%}   & \textbf{-1.56\%}    & \textbf{0.53\%}     \\ \hline
\multicolumn{1}{|c|}{\multirow{4}{*}{\textbf{\#Pin=20}}}  & \textbf{10\%}    & \textbf{-1.46\%}  & \textbf{-1.53\%}  & \textbf{-1.83\%}   & \textbf{-3.40\%}   & \textbf{-3.50\%}    & \textbf{-4.47\%}    \\
\multicolumn{1}{|c|}{}                                    & \textbf{30\%}    & \textbf{-0.27\%}  & \textbf{-0.89\%}  & \textbf{-1.73\%}   & \textbf{-2.69\%}   & \textbf{-3.84\%}    & \textbf{-4.48\%}    \\
\multicolumn{1}{|c|}{}                                    & \textbf{50\%}    & \textbf{0.47\%}   & \textbf{0.72\%}   & \textbf{-0.29\%}   & \textbf{-2.00\%}   & \textbf{-2.96\%}    & \textbf{-3.86\%}    \\
\multicolumn{1}{|c|}{}                                    & \textbf{70\%}    & \textbf{1.20\%}   & \textbf{-1.34\%}  & \textbf{-0.18\%}   & \textbf{-1.74\%}   & \textbf{-2.30\%}    & \textbf{-0.74\%}    \\ \hline
\multicolumn{1}{|c|}{\multirow{4}{*}{\textbf{\#Pin=30}}}  & \textbf{10\%}    & \textbf{-0.13\%}  & \textbf{-0.76\%}  & \textbf{-0.98\%}   & \textbf{-1.46\%}   & \textbf{-3.37\%}    & \textbf{-3.17\%}    \\
\multicolumn{1}{|c|}{}                                    & \textbf{30\%}    & \textbf{0.37\%}   & \textbf{-0.04\%}  & \textbf{-0.50\%}   & \textbf{-1.33\%}   & \textbf{-2.91\%}    & \textbf{-2.38\%}    \\
\multicolumn{1}{|c|}{}                                    & \textbf{50\%}    & \textbf{1.58\%}   & \textbf{1.46\%}   & \textbf{0.39\%}    & \textbf{-0.98\%}   & \textbf{-2.55\%}    & \textbf{-2.69\%}    \\
\multicolumn{1}{|c|}{}                                    & \textbf{70\%}    & \textbf{1.77\%}   & \textbf{2.68\%}   & \textbf{2.14\%}    & \textbf{-0.01\%}   & \textbf{-3.53\%}    & \textbf{-3.58\%}    \\ \hline
\multicolumn{1}{|c|}{\multirow{4}{*}{\textbf{\#Pin=50}}}  & \textbf{10\%}    & \textbf{-1.29\%}  & \textbf{-1.02\%}  & \textbf{-1.36\%}   & \textbf{-2.29\%}   & \textbf{-3.18\%}    & \textbf{-3.47\%}    \\
\multicolumn{1}{|c|}{}                                    & \textbf{30\%}    & \textbf{-0.85\%}  & \textbf{-0.22\%}  & \textbf{-0.19\%}   & \textbf{-2.08\%}   & \textbf{-3.21\%}    & \textbf{-3.79\%}    \\
\multicolumn{1}{|c|}{}                                    & \textbf{50\%}    & \textbf{-0.61\%}  & \textbf{0.72\%}   & \textbf{-0.11\%}   & \textbf{-1.64\%}   & \textbf{-2.75\%}    & \textbf{-2.99\%}    \\
\multicolumn{1}{|c|}{}                                    & \textbf{70\%}    & \textbf{-0.22\%}  & \textbf{1.89\%}   & \textbf{1.59\%}    & \textbf{-0.66\%}   & \textbf{-3.31\%}    & \textbf{-1.77\%}    \\ \hline
\multicolumn{1}{|c|}{\multirow{4}{*}{\textbf{\#Pin=100}}} & \textbf{10\%}    & \textbf{-1.02\%}  & \textbf{-1.01\%}  & \textbf{-1.16\%}   & \textbf{-1.59\%}   & \textbf{-2.13\%}    & \textbf{-2.51\%}    \\
\multicolumn{1}{|c|}{}                                    & \textbf{30\%}    & \textbf{-0.65\%}  & \textbf{-0.40\%}  & \textbf{-0.27\%}   & \textbf{-1.28\%}   & \textbf{-2.23\%}    & \textbf{-2.62\%}    \\
\multicolumn{1}{|c|}{}                                    & \textbf{50\%}    & \textbf{-0.87\%}  & \textbf{0.57\%}   & \textbf{0.86\%}    & \textbf{-0.43\%}   & \textbf{-1.66\%}    & \textbf{-2.60\%}    \\
\multicolumn{1}{|c|}{}                                    & \textbf{70\%}    & \textbf{-1.39\%}  & \textbf{1.29\%}   & \textbf{1.51\%}    & \textbf{-0.13\%}   & \textbf{-1.37\%}    & \textbf{-1.70\%}    \\ \hline
\multicolumn{1}{|c|}{\multirow{4}{*}{\textbf{\#Pin=200}}} & \textbf{10\%}    & \textbf{-1.15\%}  & \textbf{-0.98\%}  & \textbf{-1.02\%}   & \textbf{-1.16\%}   & \textbf{-1.56\%}    & \textbf{-1.85\%}    \\
\multicolumn{1}{|c|}{}                                    & \textbf{30\%}    & \textbf{-1.15\%}  & \textbf{-0.23\%}  & \textbf{-0.10\%}   & \textbf{-0.51\%}   & \textbf{-1.21\%}    & \textbf{-1.98\%}    \\
\multicolumn{1}{|c|}{}                                    & \textbf{50\%}    & \textbf{-1.21\%}  & \textbf{-0.18\%}  & \textbf{0.35\%}    & \textbf{0.23\%}    & \textbf{-0.63\%}    & \textbf{-1.29\%}    \\
\multicolumn{1}{|c|}{}                                    & \textbf{70\%}    & \textbf{-0.91\%}  & \textbf{-0.05\%}  & \textbf{1.03\%}    & \textbf{1.01\%}    & \textbf{0.06\%}     & \textbf{-0.92\%}    \\ \hline

\end{tabular}

}
\vspace{-2ex}
\end{table}

The OARSMT algorithm is first tested and compared. During the OARSMT algorithm comparisons, all obstacles in the layout are considered, rather than only considering those in the net's bounding box as the initial routing flow. In this case, there is no violation in the experiments of the OARSMT algorithm itself.

The wirelength difference and speedup of the proposed algorithm are shown in Table~\ref{tab:comp}, compared to previously advanced work~\cite{Lin18}. In Table~\ref{tab:comp}, each row represents a setting of the number of pins and each column represents a setting of the number of obstacles. Each group contains four runtime results corresponding to different obstacle densities. 
One can notice that the proposed algorithm has $\sim$10x - 2700x speedup from small cases to large cases.

The absolute values of the runtime for the proposed algorithm are also tested, and they are 0.63 ms for small cases and only 0.19 s for large cases. This not only shows its high speed, but also proves the scalability. 

Besides, Table~\ref{tab:comp} also shows the wirelength difference, and negative results mean that the proposed algorithm has a smaller wirelength. The wirelength difference ranges from $-$5.76\% to 2.68\%.
It is shown that the proposed algorithm can also outperform work~\cite{Lin18} in the wirelength for various cases, and the wirelength improvement is 1.14\% on average. 
For relatively large designs, the proposed algorithm shows $>$3\% wirelength improvement. 
The experiments indicate that the proposed algorithm has better wirelength for randomized testcases compared with the previous work, but also achieves orders of magnitude shorter runtime.

\begin{figure}[H]
	\centering
	\includegraphics[width=0.85\columnwidth]{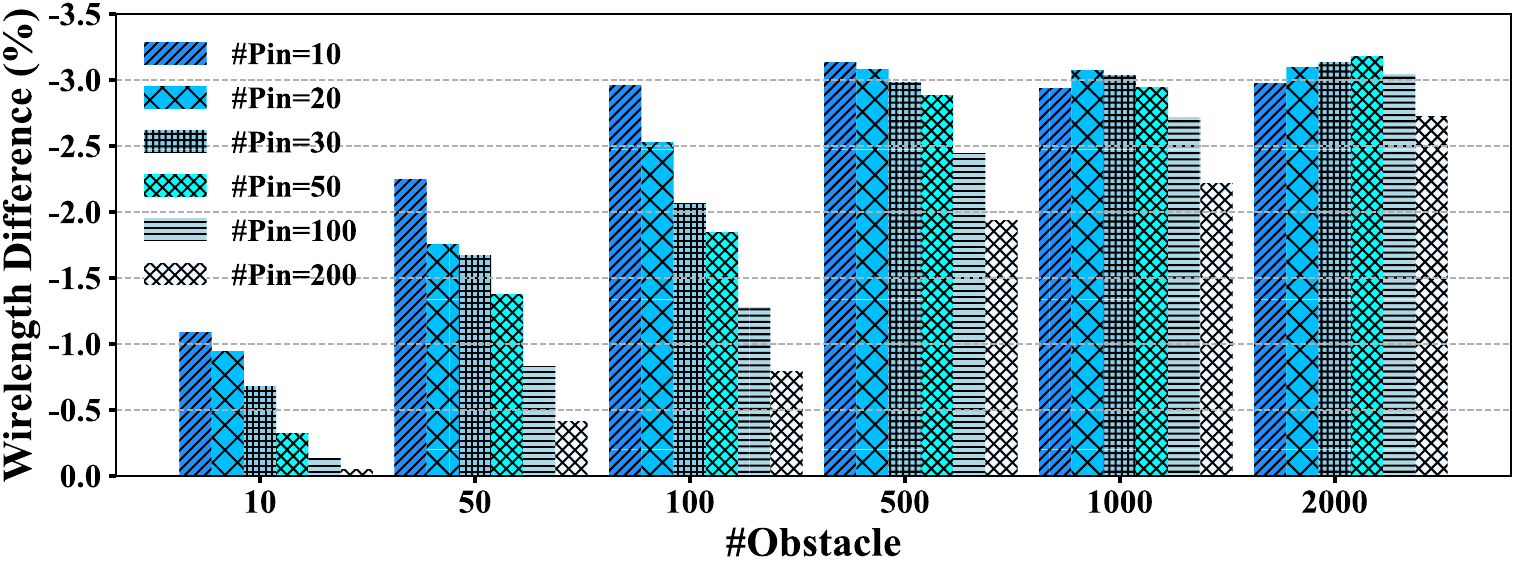}
	\caption{Wirelength improvement for rule enhancements}
	\label{fig:runtime1}
	\vspace{-2ex}
\end{figure}

Moreover, we evaluate the wirelength improvement of the enhanced rules as shown in Figure~\ref{fig:runtime1}, which shows the differences between the runs with and without the enhanced rules. The algorithm with rule enhancements shows an average -1.88\% wirelength improvement. For the cases with fewer pins (\#pins$\leq$30) or a large number of obstacles (\#obstacles$\geq$500), when the obstacles that each edge passes become complicated, it shows a bigger -2.22\% and -2.69\% wirelength improvement, respectively. The results prove that the rule enhancements, by exploring more design space, provide more opportunities to avoid local optimal solutions and achieve better wirelength.

Finally, we briefly report the wirelength and runtime comparisons with previous 2D OARSMT work~\cite{Lin08}. 
The speedup of the proposed work compared to work~\cite{Lin08} ranges from $\sim$11x - 7374x, with 1306x on average. Meanwhile, 
The wirelength difference ranges from -10.04\% to -0.47\%. The average wirelength improvement among all cases is 5.03\%.
The previous work~\cite{Lin08} results in a worse wirelength because it trims some optimized Steiner nodes when building an initial tree structure at the early stages. This again indicates the advantages of the proposed algorithm on both quality and speed.

\subsection{OARSMT Comparisons on Standard Benchmarks}

\begin{table}[!ht]
\centering
\caption{The comparisons on the standard benchmarks~\cite{oarsmtcase}.}
\label{tab:bench}

\resizebox{0.98\linewidth}{!}{

\scriptsize  
\begin{tabular}{|c|c|cccc|cccc|}
\hline
\multirow{3}{0.7cm}{\centering \textbf{Test\\cases}} & \multirow{3}{*}{\textbf{\#Pin/\#Obs}} & \multicolumn{4}{c|}{\textbf{Wirelength}} & \multicolumn{4}{c|}{\textbf{Runtime (s)}} \\ \cline{3-10} 
 &   & \multirow{2}{*}{\textbf{\cite{Lin08}}} & \textbf{GSLS} & \textbf{(Lin et al.)} & \multirow{2}{*}{\textbf{Ours}} & \multirow{2}{*}{\textbf{\cite{Lin08}}} & \textbf{GSLS} & \textbf{(Lin et al.)} & \multirow{2}{*}{\textbf{Ours}} \\ 
&   &  & \textbf{\cite{Zhang23}} & \textbf{\cite{Lin18}} &  &  & \textbf{\cite{Zhang23}} & \textbf{\cite{Lin18}} & \\ \hline
\textbf{IND1} & \textbf{10/32} & \textbf{0.6k} & \textbf{0.6k} & \textbf{0.6k} & \textbf{0.6k} & \textbf{0.08} & \textbf{0.01} & \textbf{0.01} & \textbf{0.0002} \\
\textbf{IND2} & \textbf{10/43} & \textbf{10.7k} & \textbf{9.5k} & \textbf{9.2k} & \textbf{9.7k} & \textbf{0.1} & \textbf{0.01} & \textbf{0.01} & \textbf{0.0041} \\
\textbf{IND3} & \textbf{10/50} & \textbf{0.7k} & \textbf{0.6k} & \textbf{0.6k} & \textbf{0.6k} & \textbf{0.18} & \textbf{0.01} & \textbf{0.01} & \textbf{0.0012} \\
\textbf{IND4} & \textbf{25/79} & \textbf{1.2k} & \textbf{1.1k} & \textbf{1.1k} & \textbf{1.2k} & \textbf{0.51} & \textbf{0.84} & \textbf{0.06} & \textbf{0.0042} \\
\textbf{IND5} & \textbf{33/71} & \textbf{1.4k} & \textbf{1.3k} & \textbf{1.4k} & \textbf{1.4k} & \textbf{0.41} & \textbf{0.57} & \textbf{0.02} & \textbf{0.0063} \\
\textbf{RC01} & \textbf{10/10} & \textbf{28.3k} & \textbf{26.0k} & \textbf{25.5k} & \textbf{26.1k} & \textbf{0.02} & \textbf{0.01} & \textbf{0.02} & \textbf{0.0070} \\
\textbf{RC02} & \textbf{30/10} & \textbf{44.5k} & \textbf{41.4k} & \textbf{41.0k} & \textbf{44.8k} & \textbf{0.03} & \textbf{0.04} & \textbf{0.04} & \textbf{0.0041} \\
\textbf{RC03} & \textbf{50/10} & \textbf{58.6k} & \textbf{54.2k} & \textbf{52.7k} & \textbf{60.7k} & \textbf{0.05} & \textbf{0.45} & \textbf{0.06} & \textbf{0.0089} \\
\textbf{RC04} & \textbf{70/10} & \textbf{63.7k} & \textbf{59.1k} & \textbf{56.2k} & \textbf{64.8k} & \textbf{0.08} & \textbf{0.48} & \textbf{0.11} & \textbf{0.0061} \\
\textbf{RC05} & \textbf{100/10} & \textbf{79.6k} & \textbf{74.1k} & \textbf{73.8k} & \textbf{78.9k} & \textbf{0.12} & \textbf{8.33} & \textbf{0.21} & \textbf{0.0065} \\
\textbf{RC06} & \textbf{100/500} & \textbf{86.1k} & \textbf{79.7k} & \textbf{78.7k} & \textbf{83.0k} & \textbf{27.14} & \textbf{462.88} & \textbf{7.57} & \textbf{0.0637} \\
\textbf{RC07} & \textbf{200/500} & \textbf{117.0k} & \textbf{108.8k} & \textbf{107.9k} & \textbf{113.7k} & \textbf{35.11} & \textbf{611.42} & \textbf{10.28} & \textbf{0.0623} \\
\textbf{RC08} & \textbf{200/800} & \textbf{122.4k} & \textbf{112.6k} & \textbf{111.3k} & \textbf{120.4k} & \textbf{94.38} & \textbf{475.22} & \textbf{34.34} & \textbf{0.1550} \\
\textbf{RC09} & \textbf{200/1000} & \textbf{120.4k} & \textbf{111.1k} & \textbf{110.2k} & \textbf{118.0k} & \textbf{151.69} & \textbf{604.82} & \textbf{48.18} & \textbf{0.1865} \\
\textbf{RC10} & \textbf{500/100} & \textbf{178.1k} & \textbf{164.4k} & \textbf{168.4k} & \textbf{169.0k} & \textbf{7.51} & \textbf{739.25} & \textbf{38.2} & \textbf{0.0540} \\
\textbf{RC11} & \textbf{1000/100} & \textbf{250.4k} & \textbf{232.5k} & \textbf{242.3k} & \textbf{234.3k} & \textbf{30.35} & \textbf{3260.22} & \textbf{23.57} & \textbf{0.2117} \\
\textbf{RC12} & \textbf{1000/10000} & \textbf{——} & \textbf{747.5k} & \textbf{——} & \textbf{749.7k} & \textbf{——} & \textbf{3398.36} & \textbf{——} & \textbf{0.2279} \\
\textbf{RT01} & \textbf{10/500} & \textbf{2.3k} & \textbf{2.1k} & \textbf{1.9k} & \textbf{2.3k} & \textbf{19.61} & \textbf{13.34} & \textbf{0.79} & \textbf{0.0247} \\
\textbf{RT02} & \textbf{50/500} & \textbf{50.3k} & \textbf{45.9k} & \textbf{45.0k} & \textbf{48.1k} & \textbf{24.29} & \textbf{23.11} & \textbf{5.69} & \textbf{0.0340} \\
\textbf{RT03} & \textbf{100/500} & \textbf{8.8k} & \textbf{8.0k} & \textbf{8.2k} & \textbf{8.3k} & \textbf{25.43} & \textbf{305.71} & \textbf{3.13} & \textbf{0.0437} \\
\textbf{RT04} & \textbf{100/1000} & \textbf{10.8k} & \textbf{9.7k} & \textbf{8.6k} & \textbf{12.1k} & \textbf{119.05} & \textbf{154.88} & \textbf{7.58} & \textbf{0.2569} \\
\textbf{RT05} & \textbf{200/2000} & \textbf{——} & \textbf{51.4k} & \textbf{45.2k} & \textbf{61.0k} & \textbf{——} & \textbf{622.37} & \textbf{82.12} & \textbf{0.4228} \\ \hline
\textbf{Total} & \textbf{} & \textbf{1236.0k} & \textbf{1142.6k} & \textbf{1144.6k} & \textbf{1197.8k} & \textbf{536.16} & \textbf{6661.58} & \textbf{179.88} & \textbf{1.1410} \\ \hline
\textbf{Norm.} & \textbf{} & \textbf{1.03} & \textbf{0.95} & \textbf{0.96} & \textbf{1.00} & \textbf{469.90} & \textbf{5838.37} & \textbf{157.65} & \textbf{1.00} \\ \hline

\end{tabular}

}

\begin{tablenotes}
\item {\textbullet} Each result of Lin et al.~\cite{Lin18} is selected by the better one between the reproduction and \cite{Lin18}'s report; The results of GSLS~\cite{Zhang23} are referred from its report; The results of work~\cite{Lin08} are from the reproduction as the benchmarks in its report are the old versions.
\item {\textbullet} The results with ``--'' cannot be obtained in neither 3 minutes nor the corresponding report. The total wirelength and runtime are only counted for the testcases that all 4 algorithms have results.
\end{tablenotes}

\end{table}

For OARSMT algorithms, besides the comprehensive randomized testcases and comparisons, we also compare the proposed algorithm on all standard benchmarks~\cite{oarsmtcase} adopted in the compared works, including the previously advanced work proposed by Lin et al.~\cite{Lin18}, the recent work GSLS~\cite{Zhang23}, and the traditional work~\cite{Lin08}. Each testcase generates one OARSMT for one net within a layout with obstacles. The results are shown in Table~\ref{tab:bench}. The speedup of our work ranges from $\sim$150x to 5800x, with limited wirelength overhead traded compared to the works~\cite{Lin18,Zhang23}, and with even better wirelength compared to the work~\cite{Lin08}. Note that for Lin et al.~\cite{Lin18}, there is no result for \textit{RC12} in either the reproduction or the original report, due to the large runtime. In contrast, the proposed algorithm only needs $\sim$0.2s for \textit{RC12}. Besides, the runtime of the proposed algorithm also shows the scalability as it increases almost linearly with increasing problem size, and this is in contrast with recent work GSLS~\cite{Zhang23}.

\begin{figure}[!hbt]
	\centering
	\includegraphics[width=0.98\columnwidth]{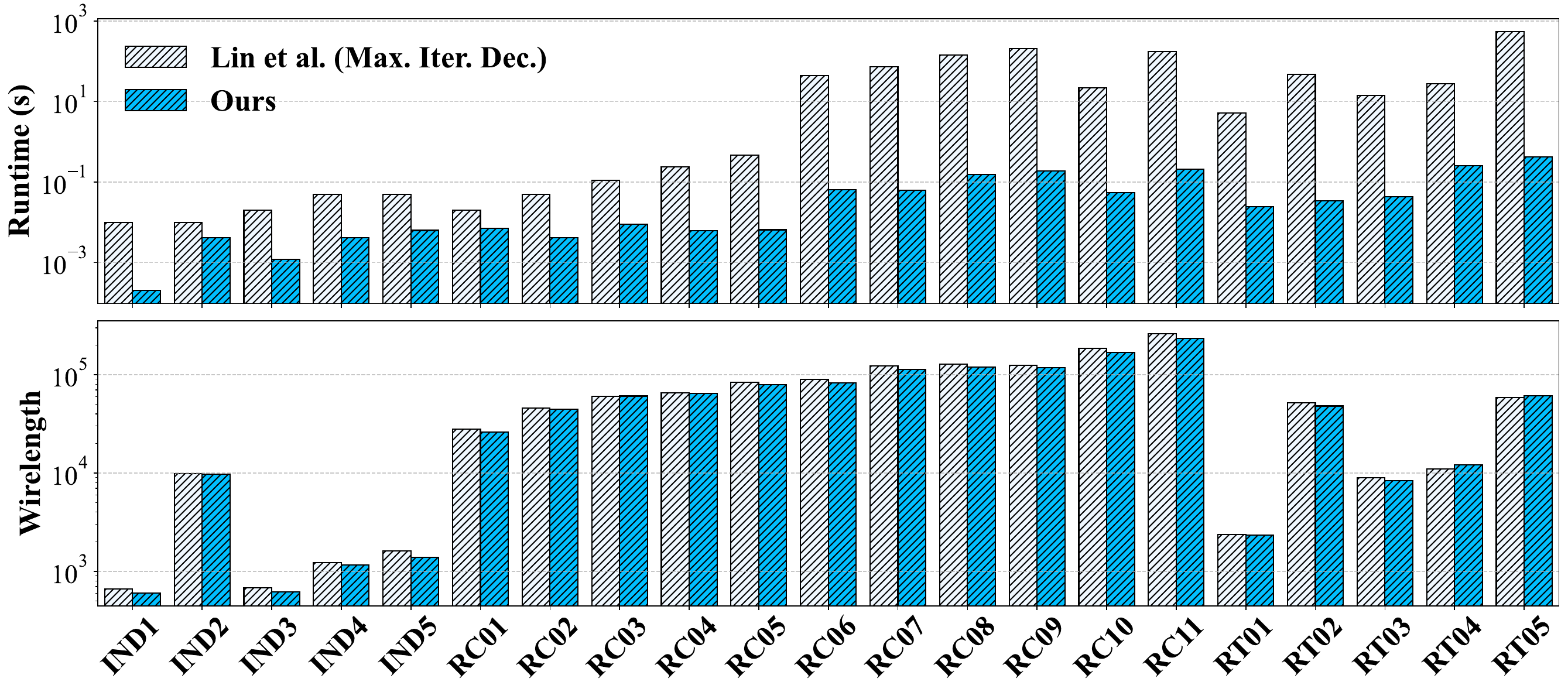}
	\caption{Performance comparison: our OARSMT vs. Lin et al.'s method with decreased iterations.}
	\label{fig:standard_OARSMT_analyse}

	\vspace{-2ex}
\end{figure}

Furthermore, we conducted an additional experiment to compare Lin et al.’s method, as shown in Figure~\ref{fig:standard_OARSMT_analyse}, based on the same industrial benchmark in Table~\ref{tab:bench}. The method proposed by Lin et al. can balance runtime and wirelength by adjusting the optimization. By default, the iteration ends when the cost does not decrease after one iteration. To ensure fair comparison and better demonstrate the superiority of our algorithm, we try to decrease its iteration to get better runtime by gradually sacrificing the wirelength, even when Lin et al.’s wirelength is already adjusted to be worse than our proposed OARSMT method, Lin et al.’s runtime is still much higher throughout the testcases. This indicates a better trade-off of our proposed rule-based OARSMT algorithm.

In complex designs with large numbers of nets, using the previous OARSMT algorithm~\cite{Lin08,Lin18} during the tree generation stage may become so computationally expensive that it exceeds the runtime of all other steps combined in a traditional routing flow. In conclusion, the experiments on standard benchmarks also demonstrate the high efficiency and scalability of the proposed algorithm. This is not only because of the theoretical lower time complexity, but also because of the simpler processing.

\subsection{Proposed Obstacle-Avoiding Global Routing Evaluations}

\begin{table}
\centering

\caption{The comparisons with other global routers.  }
\label{tab:gr_final_com}

\setlength{\tabcolsep}{3.5pt} 
\resizebox{1\linewidth}{!}{

\begin{tabular}{|c|cccc|cccc|cccc|}
\hline

\multirow{3}{2.8cm}{\centering\textbf{Benchmarks}} & \multicolumn{4}{c|}{\textbf{Normalized Wirelength}}  & \multicolumn{4}{c|}{\textbf{Normalized Via Count}} & \multicolumn{4}{c|}{\textbf{Normalized Overflow}} \\\cline{2-13}
 & \multirow{2}{*}{\textbf{CUGR 2.0}} & \textbf{CUGR 2.0} & \textbf{(Lin et al.)} & \multirow{2}{*}{\textbf{Ours}} & \multirow{2}{*}{\textbf{CUGR 2.0}} & \textbf{CUGR 2.0} & \textbf{(Lin et al.)} & \multirow{2}{*}{\textbf{Ours}}  & \multirow{2}{*}{\textbf{CUGR 2.0}} & \textbf{CUGR 2.0} & \textbf{(Lin et al.)} & \multirow{2}{*}{\textbf{Ours}}\\
 & & \textbf{Enhanced} & \textbf{\cite{Lin18}} & & & \textbf{Enhanced} & \textbf{\cite{Lin18}} & & & \textbf{Enhanced} & \textbf{\cite{Lin18}} & \\\hline

\textbf{ariane133\_51}                                                         & \textbf{1.01}                      & \textbf{1.01}     & \textbf{0.99}                      & \textbf{1.00}                  & \textbf{0.98}                      & \textbf{0.98}     & \textbf{0.98}                      & \textbf{1.00}                  & \textbf{1.33}                      & \textbf{1.33}     & \textbf{1.15}                      & \textbf{1.00}                  \\
\textbf{ariane133\_68}                                                         & \textbf{1.05}                      & \textbf{1.05}     & \textbf{0.99}                      & \textbf{1.00}                  & \textbf{0.98}                      & \textbf{0.98}     & \textbf{0.98}                      & \textbf{1.00}                  & \textbf{2.00}                      & \textbf{2.69}     & \textbf{1.10}                      & \textbf{1.00}                  \\
\textbf{nvdla}                                                                 & \textbf{1.04}                      & \textbf{1.04}     & \textbf{1.00}                      & \textbf{1.00}                  & \textbf{0.97}                      & \textbf{0.97}     & \textbf{0.98}                      & \textbf{1.00}                  & \textbf{1.38}                      & \textbf{1.40}     & \textbf{1.05}                      & \textbf{1.00}                  \\
\textbf{mempool\_tile}                                                         & \textbf{1.00}                      & \textbf{1.00}     & \textbf{0.99}                      & \textbf{1.00}                  & \textbf{0.98}                      & \textbf{0.98}     & \textbf{0.98}                      & \textbf{1.00}                  & \textbf{0.77}                      & \textbf{0.77}     & \textbf{0.77}                      & \textbf{1.00}                  \\
\textbf{bsg\_chip}                                                             & \textbf{1.02}                      & \textbf{1.02}     & \textbf{0.98}                      & \textbf{1.00}                  & \textbf{0.99}                      & \textbf{0.99}     & \textbf{0.95}                      & \textbf{1.00}                  & \textbf{2.06}                      & \textbf{2.22}     & \textbf{0.93}                      & \textbf{1.00}                  \\
\textbf{mempool\_group}                                                        & \textbf{1.00}                      & \textbf{1.03}     & \textbf{1.00}                      & \textbf{1.00}                  & \textbf{0.96}                      & \textbf{0.96}     & \textbf{0.97}                      & \textbf{1.00}                  & \textbf{0.69}                      & \textbf{1.19}     & \textbf{1.00}                      & \textbf{1.00}                  \\
\textbf{cluster}                                                               & \textbf{1.00}                      & \textbf{1.02}     & \textbf{0.99}                      & \textbf{1.00}                  & \textbf{0.99}                      & \textbf{0.99}     & \textbf{0.98}                      & \textbf{1.00}                  & \textbf{0.87}                      & \textbf{1.18}     & \textbf{0.98}                      & \textbf{1.00}                  \\
\textbf{ariane133\_51*}                                                        & \textbf{1.01}                      & \textbf{1.01}     & \textbf{0.99}                      & \textbf{1.00}                  & \textbf{0.98}                      & \textbf{0.98}     & \textbf{0.98}                      & \textbf{1.00}                  & \textbf{2.33}                      & \textbf{3.31}     & \textbf{2.11}                      & \textbf{1.00}                  \\
\textbf{ariane133\_68*}                                                        & \textbf{1.05}                      & \textbf{1.05}     & \textbf{0.99}                      & \textbf{1.00}                  & \textbf{0.98}                      & \textbf{0.98}     & \textbf{0.98}                      & \textbf{1.00}                  & \textbf{2.09}                      & \textbf{2.87}     & \textbf{1.06}                      & \textbf{1.00}                  \\
\textbf{nvdla*}                                                                & \textbf{1.04}                      & \textbf{1.04}     & \textbf{1.00}                      & \textbf{1.00}                  & \textbf{0.97}                      & \textbf{0.97}     & \textbf{0.98}                      & \textbf{1.00}                  & \textbf{1.31}                      & \textbf{1.33}     & \textbf{1.10}                      & \textbf{1.00}                  \\
\textbf{mempool\_tile*}                                                        & \textbf{1.00}                      & \textbf{1.00}     & \textbf{0.99}                      & \textbf{1.00}                  & \textbf{0.98}                      & \textbf{0.98}     & \textbf{0.98}                      & \textbf{1.00}                  & \textbf{1.06}                      & \textbf{1.06}     & \textbf{0.81}                      & \textbf{1.00}                  \\
\textbf{bsg\_chip*}                                                            & \textbf{1.02}                      & \textbf{1.02}     & \textbf{0.98}                      & \textbf{1.00}                  & \textbf{0.99}                      & \textbf{0.99}     & \textbf{0.95}                      & \textbf{1.00}                  & \textbf{1.97}                      & \textbf{2.26}     & \textbf{0.91}                      & \textbf{1.00}                  \\
\textbf{mempool\_group*}                                                       & \textbf{1.00}                      & \textbf{1.03}     & \textbf{0.99}                      & \textbf{1.00}                  & \textbf{0.96}                      & \textbf{0.96}     & \textbf{0.97}                      & \textbf{1.00}                  & \textbf{0.69}                      & \textbf{1.19}     & \textbf{0.99}                      & \textbf{1.00}                  \\
\textbf{cluster*}                                                              & \textbf{1.00}                      & \textbf{1.02}     & \textbf{0.99}                      & \textbf{1.00}                  & \textbf{0.99}                      & \textbf{0.99}     & \textbf{0.98}                      & \textbf{1.00}                  & \textbf{0.86}                      & \textbf{1.18}     & \textbf{0.98}                      & \textbf{1.00}                  \\ \hline
\textbf{Average}                                                               & \textbf{1.02}                      & \textbf{1.02}     & \textbf{0.99}                      & \textbf{1.00}                  & \textbf{0.98}                      & \textbf{0.98}     & \textbf{0.97}                      & \textbf{1.00}                  & \textbf{1.39}                      & \textbf{1.71}     & \textbf{1.07}                      & \textbf{1.00}                  \\ \hline
\end{tabular}
}
\setlength{\tabcolsep}{3.5pt} 
\resizebox{1\linewidth}{!}{
\large
\begin{tabular}{|c|cccc|cccc|}
\hline

\multirow{3}{3.2cm}{\centering\textbf{Benchmarks}} & \multicolumn{4}{c|}{\multirow{1}{10cm}{\centering \textbf{Obstacle Violation}}}  & \multicolumn{4}{c|}{\multirow{1}{10cm}{\centering \textbf{Normalized Total Runtime}}} \\\cline{2-9}
 & \multirow{2}{2.5cm}{\centering\textbf{CUGR 2.0}} & \multirow{1}{2.5cm}{\centering\textbf{CUGR 2.0}} & \multirow{1}{2.5cm}{\centering\textbf{(Lin et al.)}} & \multirow{2}{2.5cm}{\centering\textbf{Ours}} & \multirow{2}{2.5cm}{\centering\textbf{CUGR 2.0}} & \multirow{1}{2.5cm}{\centering\textbf{CUGR 2.0}} & \multirow{1}{2.5cm}{\centering\textbf{(Lin et al.)}} & \multirow{2}{2.5cm}{\centering\textbf{Ours}}\\
 & & \textbf{Enhanced} & \textbf{\cite{Lin18}} & & & \textbf{Enhanced} & \textbf{\cite{Lin18}} & \\\hline
\textbf{ariane133\_51} & \textbf{0.2k} & \textbf{0.2k} & \textbf{0} & \textbf{0} & \textbf{0.80} & \textbf{0.79} & \textbf{7.33} & \textbf{1.00} \\
\textbf{ariane133\_68} & \textbf{6.3k} & \textbf{5.7k} & \textbf{0} & \textbf{0} & \textbf{0.90} & \textbf{0.91} & \textbf{8.91} & \textbf{1.00} \\
\textbf{nvdla} & \textbf{0.4k} & \textbf{0.2k} & \textbf{0} & \textbf{0} & \textbf{0.88} & \textbf{0.89} & \textbf{5.81} & \textbf{1.00} \\
\textbf{mempool\_tile} & \textbf{0.5k} & \textbf{0.5k} & \textbf{0} & \textbf{0} & \textbf{0.84} & \textbf{0.83} & \textbf{23.72} & \textbf{1.00} \\
\textbf{bsg\_chip} & \textbf{27.1k} & \textbf{25.1k} & \textbf{0} & \textbf{0} & \textbf{0.81} & \textbf{0.78} & \textbf{10.78} & \textbf{1.00} \\
\textbf{mempool\_group} & \textbf{1777.7k} & \textbf{3.9k} & \textbf{0} & \textbf{0} & \textbf{0.71} & \textbf{0.72} & \textbf{2.05} & \textbf{1.00} \\
\textbf{cluster} & \textbf{2531.8k} & \textbf{9.5k} & \textbf{0} & \textbf{0} & \textbf{0.83} & \textbf{0.82} & \textbf{1.71} & \textbf{1.00} \\
\textbf{ariane133\_51*} & \textbf{0.2k} & \textbf{0.2k} & \textbf{0} & \textbf{0} & \textbf{0.79} & \textbf{1.05} & \textbf{8.61} & \textbf{1.00} \\
\textbf{ariane133\_68*} & \textbf{7.9k} & \textbf{7.1k} & \textbf{0} & \textbf{0} & \textbf{0.78} & \textbf{0.87} & \textbf{11.34} & \textbf{1.00} \\
\textbf{nvdla*} & \textbf{0.4k} & \textbf{0.2k} & \textbf{0} & \textbf{0} & \textbf{0.77} & \textbf{0.89} & \textbf{5.75} & \textbf{1.00} \\
\textbf{mempool\_tile*} & \textbf{0.5k} & \textbf{0.5k} & \textbf{0} & \textbf{0} & \textbf{0.75} & \textbf{0.95} & \textbf{28.43} & \textbf{1.00} \\
\textbf{bsg\_chip*} & \textbf{25.1k} & \textbf{22.8k} & \textbf{0} & \textbf{0} & \textbf{0.90} & \textbf{1.07} & \textbf{11.00} & \textbf{1.00} \\
\textbf{mempool\_group*} & \textbf{1778.1k} & \textbf{3.8k} & \textbf{0} & \textbf{0} & \textbf{0.70} & \textbf{0.77} & \textbf{1.96} & \textbf{1.00} \\
\textbf{cluster*} & \textbf{2559.9k} & \textbf{11.4k} & \textbf{0} & \textbf{0} & \textbf{0.81} & \textbf{0.79} & \textbf{1.53} & \textbf{1.00} \\ \hline
\textbf{Average} & \textbf{622.57k} & \textbf{6.51k} & \textbf{0} & \textbf{0} & \textbf{0.81} & \textbf{0.87} & \textbf{9.21} & \textbf{1.00} \\ \hline

\end{tabular}

}

\begin{tablenotes}
\footnotesize
\item {\textbullet} The Lin et al. is our proposed obstacle-avoiding global routing flow with the previously advanced OARSMT~\cite{Lin18}.
\item {\textbullet} In the benchmark, cases without ``*'' represent obstacles that only include macro obstacles, while those with ``*'' include both macro obstacles and an additional 60 randomly generated obstacles.

\end{tablenotes}
\vspace{-4ex}
\end{table}

As shown in Table~\ref{tab:gr_final_com}, we compare the proposed global router (``Ours'') with CUGR 2.0~\cite{cugr2}, enhanced CUGR 2.0, and our global routing flow integrated with the previously advanced OARSMT~\cite{Lin18} method (``Lin et al.''). 
The enhanced CUGR 2.0 is CUGR 2.0 with the increased edge cost for obstacles during maze routing. The cost is large enough to effectively make the maze routing avoid the obstacles in the sparse graphs if possible. We separately compare wirelength, via count, obstacle violation cost, overflow cost, and runtime. Note that the overflow cost does not include the obstacle violation cost.

In Table~\ref{tab:gr_final_com}, CUGR 2.0 performs poorly in terms of obstacle violations. Although the enhanced CUGR 2.0 has shown significant improvement, it still fails to effectively eliminate obstacle violations. In contrast, the proposed global routing flow (``Ours'') can reduce the final obstacle violation to zero, which proves its effectiveness. Additionally, ``Ours'' demonstrates reductions in wirelength and overflow in most test cases when compared to both the original and enhanced CUGR 2.0, achieving an average improvement of 1.96\% in wirelength and a 28.06\% reduction in overflow cost compared to enhanced CUGR 2.0, although with acceptable overhead in terms of via count and runtime, as more detours are required to navigate around obstacles. The effectiveness of obstacle elimination and overflow is primarily due to the challenges posed by complex obstacle distributions. In traditional algorithms, treating obstacles similarly to overflow complicates the balancing of costs in each stage. This disrupts the collaboration of stages in a traditional global routing process. 
Meanwhile, maze routing sparse grids in traditional global routers without OARSMT guidance also limit the opportunities of avoiding obstacles. 
In this case, addressing obstacle violations directly in the early stage and modifying the later stages to incorporate early guidance allows global routing stages to effectively manage obstacles. This is also supported by the insights presented in Section~\ref{sec:flowinsights}.

To further evaluate the effectiveness of the proposed complete flow, we also replace the OARSMT algorithm in the proposed flow with the previously advanced OARSMT~\cite{Lin18} (``Lin et al.'' in Table~\ref{tab:gr_final_com}). It shows that although both ``Ours'' and ``Lin et al.'' eliminate all obstacle violations in the benchmark, ``Lin et al.'' spends 9.21x  runtime on average than ``Ours'', with similar wirelength, via count, and overflow cost. This further demonstrates the efficiency of our obstacle-avoiding global routing flow with the proposed OARSMT algorithm.

\begin{figure}[!hbt]
	\centering
	\includegraphics[width=0.98\columnwidth]{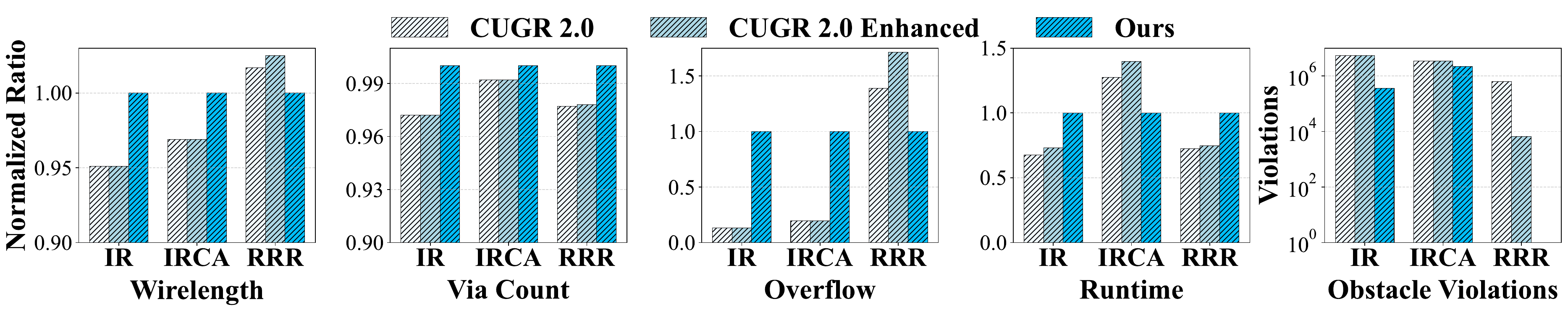}
	\caption{Comparison of CUGR 2.0, CUGR 2.0 enhanced, and proposed method across different stages in Wirelength, Via Count, Overflow, Runtime and Obstacle Violations.}
	\label{fig:each_stage_comparison}
	\vspace{-2ex}
\end{figure}

In addition, we conducted an experiment to analyze the performance of each stage. The Figure~\ref{fig:each_stage_comparison} illustrates a comparison between CUGR 2.0, enhanced CUGR 2.0, and our method ("Ours") in three stages: initial routing (IR), initial routing with congestion augmentation (IRCA), and rip-up and reroute (RRR). The comparison is based on the average values of all cases for five metrics: wirelength, via count, overflow, runtime, and obstacle violations. As shown in Figure~\ref{fig:each_stage_comparison}, our initial routing exhibits a higher overflow due to our priority consideration of avoiding obstacles. Although traditional global routing has less overflow in the early stages, they do not effectively consider obstacles. This is why, in Figure~\ref{fig:each_stage_comparison}, during initial routing, traditional routers have better wirelength and overflow, but they have worse obstacle violations. In later stages, without considering the obstacles in advance or the guidance for rerouting, transitional routers have higher difficulties handling the violations, leading to even worse wirelength and overflow, as well as remaining violations.

\subsection{Obstacle-Avoiding Global Routing Insight Analysis}

In addition to the results of the complete flow, we also analyze the insides, including runtime breakdown, the comparisons for the initial routing step, the comparisons of using different graphs during maze routing, the effectiveness of OARSMT-guided sparse maze routing, and ablation experiments isolating rule-based OARSMT and proposed maze routing separately.

\begin{figure}
	\centering
	\includegraphics[width=0.9\columnwidth]{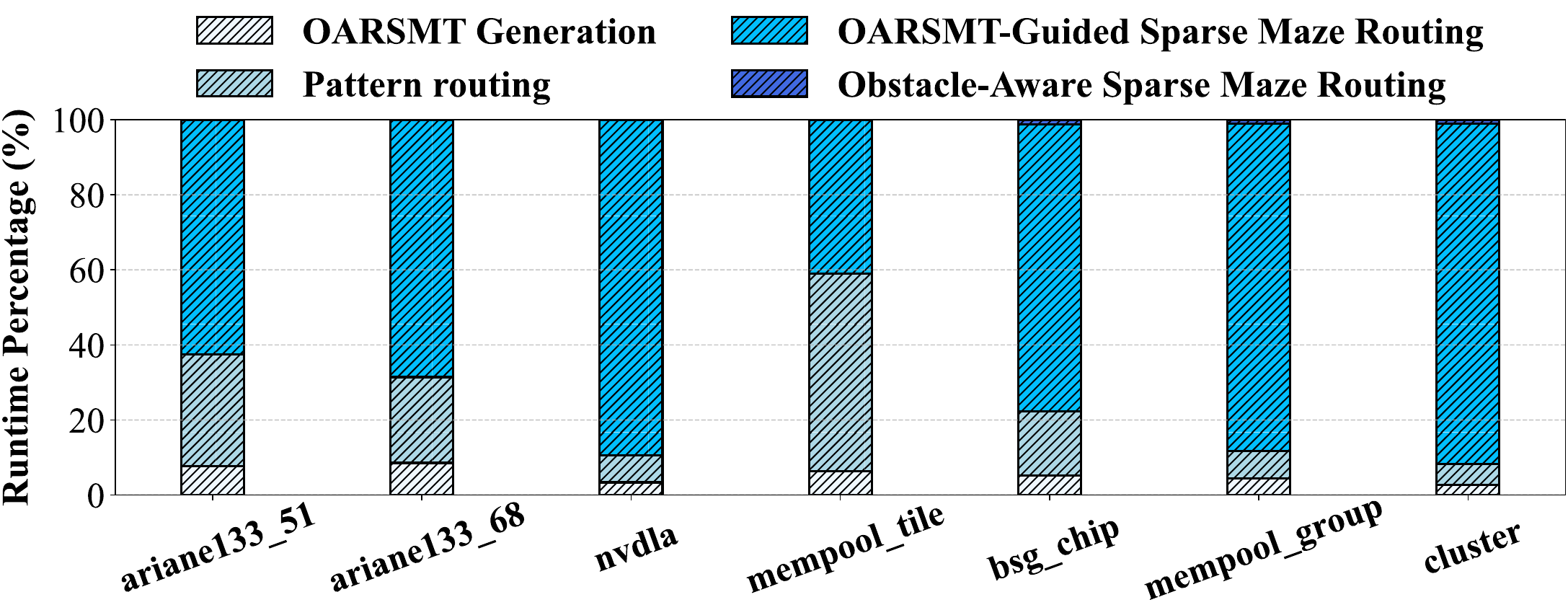}
	\caption{Runtime breakdown of the proposed flow across benchmarks, showing the percentage distribution of four stages: OARSMT generation, pattern routing, OARSMT-guided sparse maze routing, and obstacle-aware sparse maze routing.}
	\label{fig:runtime_breakdown}
        
	\vspace{-2ex}
\end{figure}

\subsubsection{Runtime Breakdown}

To further understand the runtime characteristics of the proposed global routing flow, we provide a detailed runtime breakdown analysis for each stage across different benchmarks. As shown in Figure~\ref{fig:runtime_breakdown}, similar as transitional flows, the rip-up and reroute step consumes the largest portion of the runtime. Within this step, almost all the runtime is spent by OARSMT-guided sparse maze routing. In contrast, even for large cases, the final obstacle-aware sparse maze routing consumes no more than 0.14\% of the total runtime in the worst-case scenario.
 
\subsubsection{Comparisons of Different OARSMT Algorithms for Initial Routing}

\begin{figure}[!hbt]
	\centering
	\includegraphics[width=0.9\columnwidth]{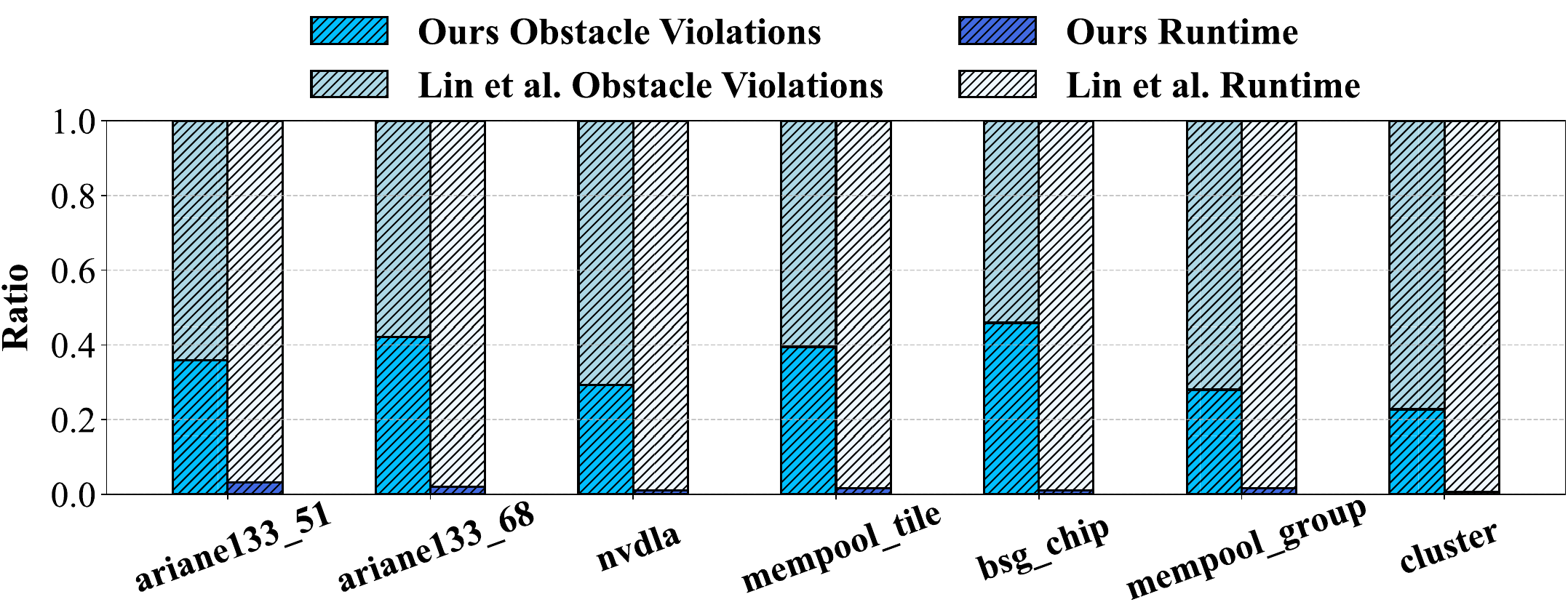}
	\caption{Obstacle violation and runtime ratio of the initial routing using the Lin et al. OARSMT and our initial routing to the CUGR 2.0.}
        
	\label{fig:initial_com}
	\vspace{-2ex}
\end{figure}

Next, Figure~\ref{fig:initial_com} illustrates the obstacle violations and runtime ratio of initial routing using the previously advanced OARSMT (``Lin et al.'') and our initial routing (``Ours'') to CUGR 2.0. Our initial routing has from 70\% to 95\% fewer obstacle violations than CUGR 2.0, from 16\% to 71\% fewer than ``Lin et al.''. ``Lin et al.'' is unable to generate effective solutions in some cases, which results in a higher obstacle violation cost during the initial routing phase compared to ``Ours''. Furthermore, ``Ours'' has $\sim$28x - 177x speedup compared with ``Lin et al.'', with an average of 149x. This again indicates the advantages of the proposed OARSMT algorithm on quality and speed towards complete designs.

\label{sec:flowinsights}

\begin{table}[!ht]
\centering
\caption{The comparison of normalized results of conventional maze routing with dense routing graphs to the proposed rip-up and reroute step, under the same initial routing flow. (Both methods have no obstacle violation)}
\label{tab:sparseMR_effective}
\resizebox{0.8\linewidth}{!}{
\begin{tabular}{|c|cccc|}
\hline
\textbf{Benchmark}      & \multicolumn{1}{c|}{\textbf{\begin{tabular}[c]{@{}c@{}}Normalized\\ Wirelength\end{tabular}}} & \multicolumn{1}{c|}{\textbf{\begin{tabular}[c]{@{}c@{}}Normalized\\ Via Count\end{tabular}}} & \multicolumn{1}{c|}{\textbf{\begin{tabular}[c]{@{}c@{}}Normalized\\ Overflow\end{tabular}}} & \textbf{\begin{tabular}[c]{@{}c@{}}Normalized\\ Runtime\end{tabular}} \\ \hline
\textbf{ariane133\_51}  & \multicolumn{1}{c|}{\textbf{0.998}}                                                           & \multicolumn{1}{c|}{\textbf{1.001}}                                                          & \multicolumn{1}{c|}{\textbf{1.16}}                                                          & \textbf{66}                                                           \\
\textbf{ariane133\_68}  & \multicolumn{1}{c|}{\textbf{0.985}}                                                           & \multicolumn{1}{c|}{\textbf{1.006}}                                                          & \multicolumn{1}{c|}{\textbf{0.83}}                                                          & \textbf{66}                                                           \\
\textbf{nvdla}          & \multicolumn{1}{c|}{\textbf{0.996}}                                                           & \multicolumn{1}{c|}{\textbf{1.038}}                                                          & \multicolumn{1}{c|}{\textbf{0.85}}                                                          & \textbf{87}                                                           \\
\textbf{mempool\_tile}  & \multicolumn{1}{c|}{\textbf{0.999}}                                                           & \multicolumn{1}{c|}{\textbf{1.005}}                                                          & \multicolumn{1}{c|}{\textbf{0.90}}                                                          & \textbf{55}                                                           \\
\textbf{bsg\_chip}      & \multicolumn{1}{c|}{\textbf{0.994}}                                                           & \multicolumn{1}{c|}{\textbf{1.012}}                                                          & \multicolumn{1}{c|}{\textbf{0.46}}                                                          & \textbf{76}                                                           \\
\textbf{mempool\_group} & \multicolumn{1}{c|}{\textbf{0.999}}                                                           & \multicolumn{1}{c|}{\textbf{1.248}}                                                          & \multicolumn{1}{c|}{\textbf{0.86}}                                                          & \textbf{77}                                                           \\ 
\textbf{cluster}        & \multicolumn{1}{c|}{-} & \multicolumn{1}{c|}{-} & \multicolumn{1}{c|}{-} & Needs $\geq$ 4 Days                                                                                                                                                                                                                                                                                                           \\ \hline
\textbf{Average}        & \multicolumn{1}{c|}{\textbf{0.995}}                                                           & \multicolumn{1}{c|}{\textbf{1.052}}                                                          & \multicolumn{1}{c|}{\textbf{0.84}}                                                          & \textbf{71}                                                           \\ \hline
\end{tabular}
}
\end{table}

\subsubsection{Comparisons of Using Different Graphs during Maze Routing}

Then, the comparisons between different rip-up and reroute approaches are analyzed. Table~\ref{tab:sparseMR_effective} shows the normalized metrics for conventional maze routing with dense routing graphs, compared to the proposed rip-up and reroute step, which includes OARSMT-guided sparse maze routing and obstacle-aware sparse maze routing. The initial routing steps both apply the proposed algorithms.
It can be noticed that, although using dense routing graphs in conventional maze routing can have better wirelength and overflow, it results in an average 71x increase in running time.
For the second-largest benchmark \textit{mempool\_group}, the proposed method requires approximately only one hour of runtime, whereas using conventional maze routing takes over 3 days. For the largest case \textit{clauster}, the conventional maze routing cannot be finished in 4 days.

Due to the large runtime, particularly for complex cases, using dense graphs for maze routing is impractical, and the proposed approach is efficient.

\subsubsection{Effectiveness Analysis of OARSMT-Guided Sparse Maze Routing}

\begin{figure}[!hbt]
	\centering
	\includegraphics[width=0.95\columnwidth]{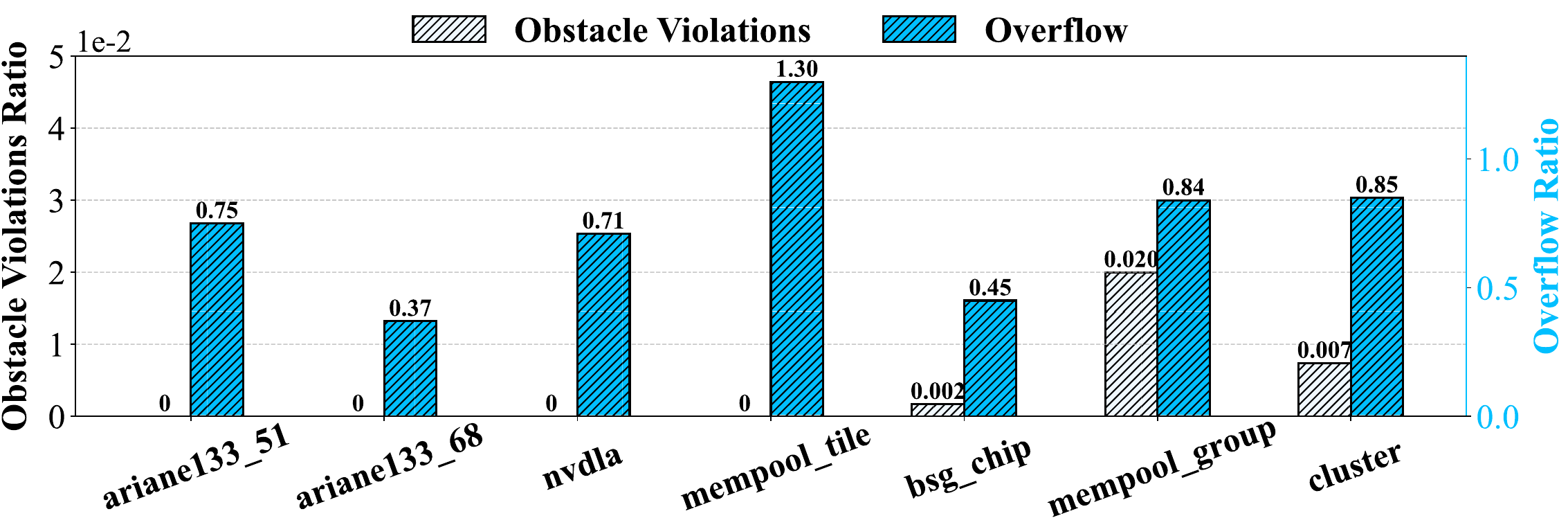}
	\caption{Obstacle violation and overflow ratio of the proposed flow after OARSMT-guided sparse maze routing to the enhanced CUGR 2.0.}
	\label{fig:guided_com}
	\vspace{-2ex}
\end{figure}

Furthermore, we analyzed the effectiveness of OARSMT-guided sparse maze routing in obstacle violations and overflow. As shown in Figure~\ref{fig:guided_com}, it illustrates obstacle violations and the overflow ratio of our obstacle-avoiding global routing flow without obstacle-aware sparse maze routing compared to the enhanced CUGR 2.0. The proposed OARSMT-guided sparse maze routing can completely eliminate obstacle violations in 4 cases. For the remaining 3 larger cases, although they are also almost cleared, their violations are significantly less than the enhanced CUGR 2.0. Except for \textit{mempool\_tile}, our OARSMT-guided sparse maze routing outperforms the enhanced CUGR2.0 in all other cases, with the best improvement reaching 58.7\% and the average improvement being 15.3\%.
This shows that using the proposed OARSMT algorithm to consider obstacles in the early stage and provide guidance for maze routing, the efficiency of avoiding the obstacles in for rip-up and reroute can be improved. 

\subsubsection{Ablation Experiments}

\begin{figure}[!hbt]
	\centering
	\includegraphics[width=0.98\columnwidth]{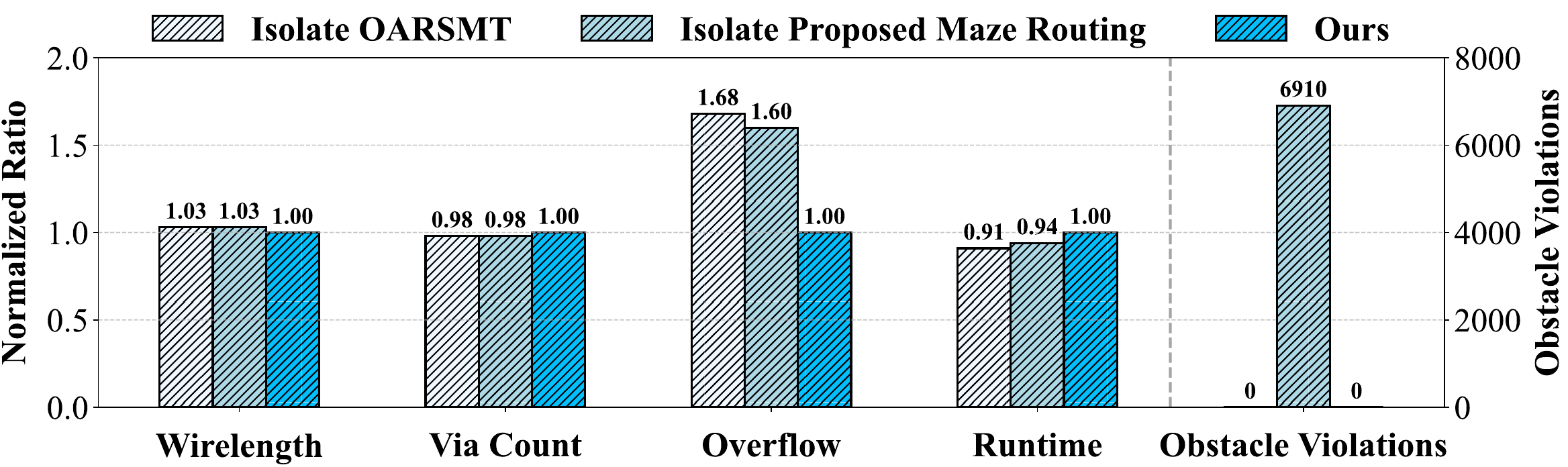}
	\caption{Ablation study on routing performance.}
	\label{fig:ablation_study}
	\vspace{-2ex}
\end{figure}

Finally, we conduct an ablation study by isolating the rule-based OARSMT and our proposed maze routing, respectively. We calculated the average values of five metrics across all testcases for three methods: wirelength, via count, overflow, runtime, and obstacle violations. All metrics except obstacle violations are normalized ratios relative to our method. As shown in Figure~\ref{fig:ablation_study}, 
the performance of using only the rule-based OARSMT method is worse than ``Ours''. Although the proposed OARSMT algorithm can handle most obstacles during tree generation, without the collaboration of the proposed maze routing, conventional sparse maze routing cannot leverage the guidance from our OARSMT structures, and thus, it cannot well handle the OARSMTs by its routing graph, leaving residual violations.
On the other hand, only using the proposed maze routing without OARSMT generation also failed to provide the best results. As there is no guidance from OARSMT in the early stages, the obstacles are intensively considered during the maze routing step, where there are fewer opportunities to adjust the routing compared with the early steps. This results in higher overflow. 
This shows that both the OARSMT and the proposed maze routing are indispensable components, and applying them together can efficiently and synergistically handle the obstacles without scarifying much wirelength and via count.

\subsection{Discussions of Optimality and Legality of the Proposed OARSMT Algorithm}
\label{sec:ruleprove}

In this section, we discuss the optimality and legality of the proposed OARSMT algorithm in Section~\ref{sec:rulebasedoarsmt} based on the experimental results.

For optimality, the proposed OARSMT algorithm is designed to support efficient obstacle-avoiding routing in practical scenarios. As the proposed algorithm spends polynomial time, it does not guarantee optimal solutions under the NP-hard nature of the OARSMT problem. However, the proposed algorithm is still able to provide slightly better solutions compared to previous works with higher efficiency, which is important for modern large-scale designs.

For legality, proving the proposed OARSMT algorithm can provide legal solutions for all cases mathematically is challenging. However, extensive experimental results (Tables~\ref{tab:comp} and~\ref{tab:bench}) show that the proposed algorithm successfully avoids obstacles for all nets in the tested benchmarks, including the industrial testcases. This proves the effectiveness in practice. 
Even though there might be some corner cases in theory, where the OARSMT algorithm fails to provide a valid solution, its probability is quite limited  (and it does not actually happen in the experiments). Moreover, these corner cases can also be resolved by rip-up and reroute step in the overall global routing flow. 
Note that there are obstacles violations during the initial routing step (such as Figures~\ref{fig:each_stage_comparison} and ~\ref{fig:initial_com}) in the proposed global routing flow, but these are different from the discussed theoretical corner cases. 
This is mainly due to the fact that when generating an OARSMT for a net during the initial routing, only obstacles within the net's bounding box are considered to improve the efficiency, according to Section~\ref{sec:initial_routing}.

\section{Conclusion}
\label{sec:conclusion}

This work proposes an effective obstacle-avoiding global routing flow. Leveraging the proposed rule-based OARSMT algorithm in initial routing, it significantly reduces the obstacle violations in the early stage. By proposing the OARSMT-guided sparse maze routing and obstacle-aware sparse maze routing in rip-up and reroute, the obstacle violations are effectively eliminated. Compared with the previously advanced OARSMT algorithm, with limited wirelength overhead, the proposed OARSMT algorithm achieves $\sim$10x–2700x and $\sim$150x–5800x runtime speedup on randomized testcases and standard benchmarks, respectively. Compared with the previously advanced enhanced global router CUGR 2.0, the proposed obstacle-avoiding global router successfully eliminates obstacle violations, and reduces wirelength and overflow cost on the modified ISPD 2024 benchmarks, with limited via count and runtime overhead. In addition, it achieves an average improvement of 1.96\% in wirelength and a reduction of 28.06\% in overflow cost.

\bibliographystyle{unsrt}  


\bibliography{references}
\end{document}